\documentclass[superscriptaddress,twocolumn,amsmath,amssymb,prd,floatfix,nofootinbib]{revtex4}

\usepackage{graphicx}
\usepackage{dcolumn}
\usepackage[varg]{txfonts}
\usepackage{bm}

\usepackage{color}

\newcommand*\Laplace{\mathop{}\!\mathbin\bigtriangleup}

\begin{document}

\title{The statistics of peaks of weakly non-Gaussian random
  fields: Effects of bispectrum in two- and three-dimensions}

\author{Takahiko Matsubara} \email{tmats@post.kek.jp}
\affiliation{%
  Institute of Particle and Nuclear Studies, High Energy
  Accelerator Research Organization (KEK), Oho 1-1, Tsukuba 305-0801,
  Japan}%
\affiliation{%
  The Graduate University for Advanced Studies (SOKENDAI),
  Tsukuba, Ibaraki 305-0801, Japan}%


\date{\today}

\begin{abstract}
  Analytic expressions for the statistics of peaks of random fields
  with weak non-Gaussianity are provided. Specifically, the abundance
  and spatial correlation of peaks are represented by formulas which
  can be evaluated only by virtually one-dimensional integrals. We
  assume the non-Gaussianity is weak enough such that it is
  represented by linear terms of the bispectrum. The formulas are
  formally given in $N$-dimensional space, and explicitly given in the
  case of $N=1,2,3$. Some examples of peak statistics in cosmological
  fields are calculated for the cosmic density field and weak lensing
  field, assuming the weak non-Gaussianity is induced by gravity. The
  formulas of this paper would find a fit in many applications to
  statistical analyses of cosmological fields.
\end{abstract}

\maketitle




\section{\label{sec:Intro}
  Introduction
}

The statistics of peaks of random fields have been attracting a lot of
interest for applications to cosmology. The density peaks are obvious
sites for the formation of nonlinear structures \cite{Kai84}. The
amplitude of spatial clustering of biased objects are enhanced
relative to that of density field \cite{DEFW85,DR87}. This property is
naturally expected by statistics of high-density peaks in a Gaussian
random field. Mathematical formalism to calculate statistics of peaks
in random Gaussian fields is given in seminal papers by Doroshkevich
\cite{Dor70} and Bardeen et al.~\cite{BBKS}. Statistics of peaks, such
as abundances, profiles and correlation functions, in Gaussian random
fields have been extensively studied in the literature
\cite{PH85,HS85,OPW86,Cou87,Col89,LHP89,PH90,RS95,Des08,MC19}. The
clustering of dark matter halos can be modeled by the peaks approach
under the assumption that halos form from peaks in the initial
Lagrangian density field (for a review, see Ref.~\cite{DJS18} and
references therein). Lagrangian density field is reasonably assumed to
obey Gaussian statistics, as long as the initial condition of the
density field in the Universe is Gaussian.

Most of the analytic work on the statistics of peaks assumes the
Gaussian statistics of density fields. One of the main reasons for
this assumption stems from technical limitations. It is extremely
difficult to analytically describe the statistics of peaks in
generally non-Gaussian fields, which have infinite degrees of freedom.
However, there are several reasons to consider the statistics of peaks
in non-Gaussian density fields in cosmology.

For example, the initial density field is not necessarily a Gaussian
random field, depending on generation mechanisms of the initial
fluctuations (see, e.g., Ref.~\cite{BKMR04} and references therein).
The gravitational evolution induces non-Gaussianity in the density
field (see, e.g., Ref.~\cite{BCGS02} and references therein), and
therefore, when the peaks are defined in Eulerian density field, they
are not described by the peak theory assuming Gaussian statistics of
density fields. The statistics of peaks in the weak lensing fields are
also useful in cosmology
\cite{JW99,HTY04,MSB09,MAPB10,FSL10,Yan11,Mar11,MSHS13,Liu14,Liu15,LK15a,HSKM15,LK15b,OSY15,Kac16,Pee18,Sha18,Mar18,Li19,Cou19}.
The weak lensing fields on interested scales are not Gaussian because
of the nonlinear evolution of the density field which is the source of
the weak lensing. The effects of non-Gaussianity are taken into
account only numerically in the previous analyses of the weak lensing.
Another example of the interest in peaks in non-Gaussian fields is the
application to the primordial black holes (PBHs), which is assumed to
be formed in the very early Universe \cite{ZN67,Haw71,CH74,CKSY10}.
The peaks theory of Ref.~\cite{BBKS} is applied to the formation of
PBHs \cite{GLMS04,YHGK18,GM19,DeL19,SY19,GS19}.

While deriving analytically complete expressions of the statistics of
peaks in generally non-Gaussian fields is difficult, that is possible
in some limited cases. For the peak abundance in a special type of
non-Gaussian field, chi-square field, an analytic expression can be
derived \cite{Blo16,Blo18}. A theory for the abundance of peaks in
weakly non-Gaussian fields is pioneered by Refs.~\cite{PGP09,GPP12},
which generalize the earlier work on the genus statistic and Minkowski
functionals in weakly non-Gaussian fields \cite{Mat94,Mat03}. In these
papers, the peak abundances in two and three dimensions are expanded
in Gram-Charlier series \cite{Cha67,Jus95,Mat95,Ame96}. When the
non-Gaussianity is weak, and the higher-order cumulants of the
distribution do not significantly contribute to the statistics of
peaks, one obtains an approximate expression for peak abundances by
only taking lower-order terms of the series into account. The peak
correlations in weakly non-Gaussian fields are derived \cite{DGR13},
which are applied to a local-type non-Gaussianity in the primordial
density field. Abundances and correlations of peaks in weakly
non-Gaussian field in the high-peak limit are also derived
\cite{CLM88,GW86,MLB86,ARW06}.

In this paper, we follow and extend the methods of those previous
papers for peaks in weakly non-Gaussian field, and give explicit
formulas with lowest-order non-Gaussianity in two and three
dimensions. We consider the abundances and spatial correlations of
peaks in a unified formalism which is developed by Ref.~\cite{Mat95}.
We first show a formal derivation of the peak statistics in $N$
dimensions, and then find explicit expressions for $N=1,2,3$. In
Ref.~\cite{GPP12}, the formulas for the abundance of peaks are given
in a form with multi-dimensional integrations, which should be
evaluated by a semi-Monte-Carlo integration. We find this kind of
multi-dimensional integrations reduces to lower-dimensional integrals,
which can be evaluated very fast, extending techniques developed by
Refs.~\cite{LMD16,MD16}. This paper contains a set of newly useful
formulas for statistics of peaks of weakly non-Gaussian fields, which
can be potentially applied to many problems regarding statistics of
peaks, such as the peaks in the density field of large-scale structure
and in weak lensing fields, etc.

This paper is organized as follows. In Sec.~\ref{sec:Abundance}, a
formal expression of the number density of peaks in a weakly
non-Gaussian field in an $N$-dimensional space is given, and then
analytically explicit expressions for $N=1,2,3$ are derived. In
Sec.~\ref{sec:Correlations}, formal expressions of the power spectrum
and correlation function of peaks in a weakly non-Gaussian field in an
$N$-dimensional space is given, and then analytically explicit
expressions for $N=2,3$ are derived. In Sec.~\ref{sec:GravInst}, three
examples of the possible applications to cosmology are presented,
i.e., the number density of peaks in a three-dimensional density
field, the number density of peaks in a two-dimensional weak lensing
field, and three-dimensional correlations of peaks. In these examples,
the weak non-Gaussianity is assumed to emerge from weakly nonlinear
evolutions by gravitational instability. Finally, conclusions are
given in Sec.~\ref{sec:Conclusions}.

\section{\label{sec:Abundance}
  Abundance of peaks in weakly non-Gaussian fields
}

\subsection{\label{subsec:AbunLowestGauss}
  Lowest-order non-Gaussianity
}

We generally consider a random field $f(\bm{x})$ in $N$-dimensional
space, where $\bm{x}$ is the $N$-dimensional coordinates. The field is
assumed to have a zero-mean,
\begin{equation}
  \label{eq:2-1}
  \langle f(\bm{x}) \rangle = 0,
\end{equation}
and the random field is statistically homogeneous and isotropic. We
consider expectation values of peak statistics in non-Gaussian fields.
We apply a method of Ref.~\cite{Mat95}, which provides a general way
of evaluating a given expectation value in weakly non-Gaussian fields.
The method is based on the expansion by generalized Wiener-Hermite
functionals, which is a generalization of the Edgeworth expansion of a
single variable in weakly non-Gaussian fields. This basic method is
briefly reviewed in Appendix~\ref{app:Expansion}.

In this paper, we consider the lowest-order non-Gaussianity, i.e.,
contributions from the three-point correlation at the lowest order,
assuming the higher-order correlations are small enough. In
cosmological fields, higher-order correlations frequently obey the
so-called hierarchical ordering, in which $n$-point correlation
function $\xi^{(n)}$ is of order $\mathcal{O}(\xi^{n-1})$, where
$\xi=\xi^{(2)}$ is the two-point correlation function. In this case,
the non-Gaussianity is weak when the two-point correlation $\xi$ is
small enough.

Having such a case in our mind, we consider only the linear
contribution of the three-point correlation function, or the
bispectrum in Fourier space. The expectation value of a functional
$\mathcal{F}[f]$ is given by Eq.~(\ref{eq:a-16}). When we take into
account only the lowest-order non-Gaussianity, we have
\begin{multline}
  \label{eq:2-2}
  \left\langle \mathcal{F}[f] \right\rangle =
  \left\langle \mathcal{F}[f] \right\rangle_\mathrm{G}
  \\
  +
  \frac{1}{6}
  \int \frac{d^N\!k_1}{(2\pi)^N} \frac{d^N\!k_2}{(2\pi)^N}
    \frac{d^N\!k_3}{(2\pi)^N} 
  \left\langle
    \tilde{f}(\bm{k}_1) \tilde{f}(\bm{k}_2) \tilde{f}(\bm{k}_3)
  \right\rangle_\mathrm{c}
  \\
  \times
  \mathcal{G}_3(\bm{k}_1,\bm{k}_2,\bm{k}_3),
\end{multline}
where $\tilde{f}(\bm{k})$ is the Fourier transform of $f(\bm{x})$,
$\langle\cdots\rangle_\mathrm{c}$ represents the (three-point)
cumulant, and
\begin{equation}
  \label{eq:2-3}
  \mathcal{G}_{n}\left(\bm{k}_1,\ldots,\bm{k}_n\right) \equiv
  (2\pi)^{Nn}
  \left\langle
    \frac{\delta^n\mathcal{F}[f]}
    {\delta \tilde{f}((\bm{k}_1) \cdots \delta \tilde{f}(\bm{k}_n)} 
  \right\rangle_\mathrm{G},
\end{equation}
represent a Gaussian $n$-point response function, and the expectation
value $\langle\cdots\rangle_\mathrm{G}$ is taken for Gaussian
distributions with the same power spectrum of the field $f(\bm{x})$
(see Appendix~\ref{app:Expansion} for details).

Due to statistical homogeneity, the three-point cumulant has a form,
\begin{equation}
  \label{eq:2-4}
  \left\langle
    \tilde{f}(\bm{k}_1) \tilde{f}(\bm{k}_2) \tilde{f}(\bm{k}_3)
  \right\rangle_\mathrm{c} =
  (2\pi)^N\delta_\mathrm{D}^N\left(\bm{k}_1 + \bm{k}_2 + \bm{k}_3\right)
  B(\bm{k}_1,\bm{k}_2,\bm{k}_3),
\end{equation}
where $\delta_\mathrm{D}^N(\bm{k})$ is the $N$-dimensional Dirac's
delta function, and $B(\bm{k}_1,\bm{k}_2,\bm{k}_3)$ is the bispectrum.
Due to statistical homogeneity and isotropy, the bispectrum is a
function of only magnitudes of three wavevectors, $k_1$, $k_2$, and
$k_3$. However, we keep the vector notation in the argument of the
bispectrum. Thus Eq.~(\ref{eq:2-2}) can also be represented by
\begin{multline}
  \label{eq:2-5}
  \left\langle \mathcal{F}[f] \right\rangle =
  \mathcal{G}_0
  \\
  +
  \frac{1}{6}
  \int \frac{d^N\!k_1}{(2\pi)^N} \frac{d^N\!k_2}{(2\pi)^N}
  \frac{d^N\!k_3}{(2\pi)^N}
  (2\pi)^N\delta_\mathrm{D}^N(\bm{k}_1 + \bm{k}_2 + \bm{k}_3)
  \\
  \times
  B(\bm{k}_1,\bm{k}_2,\bm{k}_3)
  \mathcal{G}_3(\bm{k}_1,\bm{k}_2,\bm{k}_3).
\end{multline}

\subsection{\label{subsec:GaussianPDF}
  Statistics of field derivatives
}

The Eq.~(\ref{eq:2-5}) is the basic formula of the weakly non-Gaussian
expectation values of any kind. In this section, we are interested in
the peak abundance of the weakly non-Gaussian field. The peak number
density depends on spatial derivatives of the field up to the second
order, i.e., $f$, $\partial_i f$ and $\partial_i\partial_j f$. To
evaluate the Eq.~(\ref{eq:2-3}), we need the Gaussian statistics of
the peak number density.

The power spectrum $P(k)$ of the random field $f$ is defined by
\begin{equation}
  \label{eq:2-11}
  \left\langle \tilde{f}(\bm{k}) \tilde{f}(\bm{k}')
  \right\rangle_\mathrm{c} = (2\pi)^N \delta(\bm{k} + \bm{k}')
  P(k),
\end{equation}
where the appearance of the delta function is a consequence of the
statistical homogeneity, and the power spectrum is a function of only
the magnitude of the wavevector $k=|\bm{k}|$ due to the statistical
isotropy. The spectral moment $\sigma_n$ is defined by
\begin{equation}
  \label{eq:2-12}
  {\sigma_n}^2 =
  \int \frac{d^N\!k}{(2\pi)^N} k^{2n} P(k)
\end{equation}
and the normalized field variables are defined by
\begin{equation}
  \label{eq:2-13}
  \alpha \equiv \frac{f}{\sigma_0}, \quad
  \eta_i \equiv \frac{\partial_i f}{\sigma_1}, \quad
  \zeta_{ij} \equiv \frac{\partial_i \partial_j f}{\sigma_2},
\end{equation}
where $\partial_i = \partial/\partial x_i$ is the spatial derivative.

The Gaussian statistics of the field variables are completely
determined by their covariances. They are given by
\begin{align}
  &
    \left\langle \alpha^2 \right\rangle = 1, \quad
  \langle \alpha\eta_i\rangle = 0,\quad
  \langle \alpha\zeta_{ij}\rangle = -\frac{\gamma}{N}\delta_{ij},\quad
  \langle \eta_i\eta_j\rangle = \frac{1}{N}\delta_{ij},
  \nonumber
  \\
  &
    \label{eq:2-14}
  \langle\eta_i\zeta_{jk}\rangle = 0, \quad
  \langle\zeta_{ij}\zeta_{kl}\rangle =
  \frac{1}{N(N+2)}
  (\delta_{ij}\delta_{kl}+\delta_{ik}\delta_{jl}+\delta_{il}\delta_{jk}),
\end{align}
where
\begin{equation}
  \label{eq:2-15}
  \gamma \equiv \frac{{\sigma_1}^2}{\sigma_0\sigma_2}.
\end{equation}
Since the set of variables $\zeta_{ij}$ is a symmetric
tensor, only components with $i\geq j$ are independent.

We denote the set of independent variables as
\begin{equation}
  \label{eq:2-16}
  \bm{Y} = \left(\alpha,\eta_1,\ldots,\eta_N,\zeta_{11},
    \zeta_{12},\ldots,\zeta_{N-1,N},\zeta_{NN}\right)
\end{equation}
The number of components of this vector is
$N_0 \equiv 1 + N + N(N+1)/2 = (N+1)(N+2)/2$. The multivariate Gaussian
distribution function for these variables at a single point is given
by
\begin{equation}
  \label{eq:2-17}
    \mathcal{P}_\mathrm{G}(\bm{Y})
  =
  \frac{1}{\sqrt{(2\pi)^{N_0} \det M}}
  \exp\left(-\frac{1}{2}\bm{Y}^\mathrm{T} M^{-1} \bm{Y}\right),
\end{equation}
where $M_{ab} \equiv \langle X_a X_b \rangle$ is a $N_0 \times N_0$
covariance matrix given by Eq.~(\ref{eq:2-14}). It is useful to define
the rotationally invariant quantities,
\begin{align}
  \label{eq:2-18a}
  &
    \eta^2 \equiv \bm{\eta}\cdot\bm{\eta}, \quad
    J_1 \equiv -\zeta_{ii}, \quad
  \\
  \label{eq:2-18b}
  &
    J_2 \equiv \frac{N}{N-1} \tilde{\zeta}_{ij}
    \tilde{\zeta}_{ji},\quad (N\geq 2),
  \\
  \label{eq:2-18c}
  &
    J_3 \equiv \frac{N^2}{(N-1)(N-2)} \tilde{\zeta}_{ij} \tilde{\zeta}_{jk}
  \tilde{\zeta}_{ki},\quad (N\geq 3), 
\end{align}
where repeated indices are summed over and
\begin{equation}
  \label{eq:2-19}
  \tilde{\zeta}_{ij} \equiv \zeta_{ij} + \frac{1}{N} \delta_{ij} J_1,
\end{equation}
is the traceless part of $\zeta_{ij}$. The variable $J_2$ is
considered only for $N\geq 2$ and the variable $J_3$ is considered
only for $N\geq 3$. 

In terms of the rotationally
invariant variables, the multivariate Gaussian distribution of
Eq.~(\ref{eq:2-17}) is represented by \cite{PGP09,GPP12}
\begin{multline}
  \label{eq:2-20}
    \mathcal{P}_\mathrm{G}(\bm{Y})
    \propto
    \mathcal{N}(\alpha,J_1)
  \exp\left[
    - \frac{N}{2} \eta^2 - \frac{(N-1)(N+2)}{4} J_2
  \right],
\end{multline}
up to the normalization constant, where 
\begin{equation}
  \label{eq:2-21}
  \mathcal{N}(\alpha,J_1) =
  \frac{1}{2\pi\sqrt{1 - \gamma^2}}
  \exp\left[
    -\frac{\alpha^2 + {J_1}^2 - 2\gamma\alpha J_1}{2(1-\gamma^2)}
  \right]
\end{equation}
is the Gaussian joint distribution function of variables $\alpha$ and
$J_1$. 

\subsection{\label{subsec:NumberDensity}
  The number density of peaks in a weakly non-Gaussian field 
}

The number density of peaks above a threshold $f \geq \nu \sigma_0$ is
given by \cite{BBKS}
\begin{equation}
  \label{eq:2-31}
  n_\mathrm{pk}(\nu) = 
  \left(\frac{\sigma_2}{\sigma_1}\right)^N \Theta(\alpha-\nu) 
  \delta^N(\bm{\eta})\Theta(\lambda_N)|\det\zeta|,
\end{equation}
where $\Theta(x)$ is the Heaviside's step function, and $\lambda_N$ is
the smallest eigenvalue of the $N\times N$ matrix $(-\zeta_{ij})$. In
order to obtain the weakly non-Gaussian corrections of
Eq.~(\ref{eq:2-5}), the Gaussian expectation value of
Eq.~(\ref{eq:2-3}) should be evaluated for
$\mathcal{F} = n_\mathrm{pk}$. The calculation is straightforward but
somehow complicated, and the detailed derivation is given in
Appendix~\ref{app:Gresp}. The result is usefully represented by using
coefficients defined by
\begin{multline}
  \label{eq:2-32}
  G_{ijklm}(\nu)
  \equiv
  (-1)^k
  \biggl\langle
  n_\mathrm{pk}(\nu) H_{ij}(\alpha,J_1)
  \\ \times
    L^{(N/2-1)}_k\left(\frac{N}{2}\eta^2\right)
    F_{lm}(J_2,J_3)
  \biggr\rangle_\mathrm{G},
\end{multline}
where
\begin{equation}
  \label{eq:2-33}
  H_{ij}(\nu, J_1)
  = \frac{1}{\mathcal{N}(\alpha,J_1)}
  \left(-\frac{\partial}{\partial \alpha}\right)^i 
    \left(-\frac{\partial}{\partial J_1}\right)^j \mathcal{N}(\alpha,J_1),
\end{equation}
is the multivariate Hermite polynomials,
\begin{equation}
  \label{eq:2-34}
  L^{(a)}_k(x) = \frac{x^{-a} e^x}{k!} \frac{d^k}{dx^k}
  \left(x^{k+a} e^{-x}\right).
\end{equation}
is the generalized Laguerre polynomials,
\begin{multline}
  \label{eq:2-35}
  F_{lm}(J_2,J_3) \equiv
  (-1)^l {J_2}^{3m/2}
  \\
  \times
  L^{(3m + (N-2)(N+3)/4)}_l\left(\frac{(N-1)(N+2)}{4}J_2\right)
  P_m\left(\frac{J_3}{{J_2}^{3/2}}\right),
\end{multline}
and
\begin{equation}
  \label{eq:2-36}
  P_m(x) = \frac{1}{2^mm!}\frac{d^m}{dx^m} (x^2-1)^m
\end{equation}
is the Legendre polynomials\footnote{The function
  $F_{lm}(J_2,J_3)$ corresponds to the function
  $\tilde{F}_{lm}(5J_2,J_3)$ of Ref.~\cite{LMD16} and the function
  $F_{lm}(5J_2,J_3)$ of Ref.~\cite{Diz16} in three-dimensions, but the
  normalization is different. Denoting the latter function as
  $F_{lm}^\mathrm{Diz}$, they are related by
  \begin{equation}
    \nonumber
    \tilde{F}_{lm}(5J_2,J_3) =
    (5/2)^{3m/2}\sqrt{(2m+1)\Gamma(5/2)/\Gamma(l+3m+5/2)}\,F_{lm}(J_2,J_3)
  \end{equation}
  and
  \begin{equation}
    \nonumber
    F_{lm}^\mathrm{Diz}(5J_2,J_3) =
    (5/2)^{3m/2}\sqrt{\Gamma(5/2)/\Gamma(3m+5/2)}\,F_{lm}(J_2,J_3)
  \end{equation}
  when $N=3$ (a factor $s^{3m/2}$ is missing in Eq.~(2.18) of
  Ref.~\cite{Diz16}). Accordingly, the normalizations of bias
  parameters $c_{ijklm}$ defined later in this paper are different
  from these literatures for $m\ne 0$.}. We assume $m=0$ when $N=2$,
and $l=m=0$ when $N=1$. The result of $\mathcal{G}_3$ is given by
Eq.~(\ref{eq:b-11d}) in Appendix~\ref{app:Gresp}. In the case of
$N=1$, the terms of $G_{ijklm}$ with $l\ne 0$ or $m\ne 0$ should be
omitted. In the case of $N=2$, the terms with $G_{ijklm}$ with
$m\ne 0$ should be omitted. These rules always apply in the following.
Substituting Eq.~(\ref{eq:b-11d}) into Eq.~(\ref{eq:2-5}), we derive
\begin{multline}
  \label{eq:2-41}
  \bar{n}_\mathrm{pk}(\nu) \equiv
  \left\langle n_\mathrm{pk}(\nu) \right\rangle
  \\
  = G_{00000}
  +
  \frac{\sigma_0}{6}
  \Biggl[
  G_{30000} S^{(0)}
  + 4 \gamma G_{21000} S^{(1)}
  + 3 \gamma^2 G_{12000} S^{(2)}_2
  \\
  + \gamma^3 G_{03000} S^{(3)}_1
  + 4 G_{10100} S^{(1)}
  + \frac{4(N-1)}{N}\gamma G_{01100} S^{(2)}
  \\
  + 6 \gamma^2 G_{10010} \left( S^{(2)}_2 - S^{(2)}\right)
  + \frac{6}{N-1} \gamma^3 G_{01010}
  \left(NS^{(3)}_2 - S^{(3)}_1\right)
  \\
  + \frac{3(N-2)(N+2)^2}{2(N+4)}\gamma^3 G_{00001}
  \left(\frac{N+2}{3} S^{(3)}_1 - N S^{(3)}_2 \right)
  \Biggr],
\end{multline}
where
\begin{align}
  \nonumber
  &
    S^{(0)} \equiv
    \frac{\left\langle f^3 \right\rangle_\mathrm{c}}{{\sigma_0}^4},
    \quad
    S^{(1)} \equiv
    - \frac{3}{4}
    \frac{\left\langle f^2 \Laplace f\right\rangle_\mathrm{c}}
    {{\sigma_0}^2{\sigma_1}^2},
  \\
  \nonumber
  &
    S^{(2)} \equiv
    - \frac{3N}{2(N-1)}
    \frac{\left\langle (\bm{\nabla}f\cdot\bm{\nabla}f) \Laplace
    f\right\rangle_\mathrm{c}} 
    {{\sigma_1}^4},
    \quad
    S^{(2)}_2 \equiv
    \frac{\left\langle f (\Laplace f)^2 \right\rangle_\mathrm{c}} 
    {{\sigma_1}^4},
  \\
  \label{eq:2-42} 
  &
    S^{(3)}_1 \equiv
    - \frac{{\sigma_0}^2}{{\sigma_1}^6}
    \left\langle (\Laplace f)^3 \right\rangle_\mathrm{c},
    \quad
    S^{(3)}_2 \equiv
    - \frac{{\sigma_0}^2}{{\sigma_1}^6}
    \left\langle f_{ij} f_{ij}\Laplace f \right\rangle_\mathrm{c}.
\end{align}
For $N=1$, we define $S^{(2)}\equiv 0$ because we have an identity
$\langle (f')^2f''\rangle = 0$. In deriving Eq.~(\ref{eq:2-41}), we
use identities
\begin{align}
  \nonumber
  &
    \left\langle f\bm{\nabla}f \cdot\bm{\nabla} f \right\rangle_\mathrm{c}
    = \frac{2}{3} {\sigma_0}^2 {\sigma_1}^2 S^{(1)}
  \\
  \nonumber
  &
  \left\langle f f_{ij} f_{ij} \right\rangle_\mathrm{c}
    = {\sigma_1}^4 \left(S^{(2)}_2 - \frac{N-1}{N} S^{(2)}\right),
  \\
  \nonumber
  &
  \label{eq:2-43}
  \left\langle f_i f_j f_{ij} \right\rangle_\mathrm{c}
  = \frac{N-1}{3N} {\sigma_1}^4 S^{(2)},
  \\
  &
  \left\langle f_{ij} f_{jk} f_{ki} \right\rangle_\mathrm{c}
  = \frac{{\sigma_1}^6}{2{\sigma_0}^2}
  \left(S^{(3)}_1 - 3 S^{(3)}_2 \right),
\end{align}
which are shown by integrations by parts.

\subsection{\label{subsec:CalCoeff}
  Calculating coefficients
}

The remaining task is to calculate the coefficients $G_{ijklm}$ of
Eq.~(\ref{eq:2-32}). Substituting Eqs.~(\ref{eq:2-20}) and
(\ref{eq:2-31}) into Eq.~(\ref{eq:2-32}), we have
\begin{multline}
  \label{eq:2-51}
  G_{ijklm}(\nu) = N_1
  \left(\frac{\sigma_2}{\sigma_1}\right)^N
  \left(\frac{N}{2\pi}\right)^{N/2}
  X_k \int \prod_{i\leq j}d\zeta_{ij} \Theta(\lambda_N)
  |\det\zeta|
  \\ \times
  H_{i-1,j}(\nu,J_1) F_{lm}(J_2,J_3)
  \mathcal{N}(\nu,J_1)
  \\ \times
  \exp\left[-\frac{(N-1)(N+2)}{4}J_2\right],
\end{multline}
where
\begin{equation}
  \label{eq:2-52}
  X_k \equiv (-1)^k L_k^{(N/2-1)}(0)
  = \frac{(-1)^k\Gamma(k + N/2)}{\Gamma(k+1)\Gamma(N/2)},
\end{equation}
and $N_1$ is a normalization factor defined by
\begin{align}
  \label{eq:2-53}
  {N_1}^{-1}
  &\equiv
    \int d\alpha \prod_{i\leq j} d\zeta_{ij}\,
    \mathcal{N}(\alpha,J_1)
    \exp\left[-\frac{(N-1)(N+2)}{4}J_2\right]
    \nonumber \\
  &=
    \frac{1}{\sqrt{2\pi}}
    \int \prod_{i\leq j} d\zeta_{ij}\,
    \exp\left[-\frac{{J_1}^2}{2}-\frac{(N-1)(N+2)}{4}J_2\right].
\end{align}
For $i=0$, the functions $H_{-1,j}(\nu,J_1)$ are defined by
\begin{equation}
  \label{eq:2-54}
  H_{-1,j}(\nu,J_1) \equiv
  \frac{1}{\mathcal{N}(\nu,J_1)}
  \int_\nu^\infty d\alpha\,H_{0j}(\alpha,J_1)\mathcal{N}(\alpha,J_1).
\end{equation}
In deriving the Eqs.~(\ref{eq:2-51}) and (\ref{eq:2-53}), we use the
property
$\int d\alpha \mathcal{N}(\alpha,J_1) = e^{-{J_1}^2/2}/\sqrt{2\pi}$
and the fact that the Gaussian probability distribution function of
$\bm{\eta}$ is given by
$P_\mathrm{G}(\bm{\eta})d^N\eta = (N/2\pi)^N e^{-N\eta^2/2} d^N\eta$.
We change the integration variables as
\begin{equation}
  \label{eq:2-55}
  N_1 \prod_{i\leq j} d\zeta_{ij}
  = \frac{1}{\Omega_N} dx\,d^{N-1}\!W\,d\Omega_N,
\end{equation}
where $x=J_1=\lambda_1+\cdots +\lambda_N$, $d^{N-1}\!W$ represents the
volume element of the other (traceless components of) rotationally
invariant variables, and $d\Omega_N$ represents the volume element of
the rotationally variant (angular) components, and
\begin{equation}
  \label{eq:2-56}
  \Omega_N = \int d\Omega_N = \mathrm{Vol}_{\mathrm{SO}(N)}
  =\frac{2^{N-1} \pi^{(N-1)(N+2)/4}}{\prod_{n=2}^N \Gamma(n/2)},
  \quad (N\geq 2),
\end{equation}
is the volume of $N$-dimensional rotation group SO($N$).
In practice, the volume element $d^{N-1}\!W$ is obtained by
rotating the orthogonal set of coordinates to the principal axes of
$\zeta_{ij}$ to have the diagonal form $-(\lambda_1,\lambda_2,\ldots,\lambda_N)$,
ordered by $\lambda_1 \geq \lambda_2 \geq \cdots \geq \lambda_N$. 

Because of Eqs.~(\ref{eq:2-53}) and (\ref{eq:2-55}), the normalization
condition of variables $\bm{W}$ should be
\begin{equation}
  \label{eq:2-57}
  \int_D d^{N-1}\!W\,\exp\left[-\frac{(N-1)(N+2)}{4}J_2\right] = 1,
\end{equation}
where $D$ is the integration domain to satisfy the ordering
$\lambda_1 \geq \cdots \geq \lambda_N$. Thereby,
Eq.~(\ref{eq:2-51}) can be represented as
\begin{multline}
  \label{eq:2-58}
  G_{ijklm}(\nu) = \frac{1}{(2\pi)^{N/2}}
  \left(\frac{\sigma_2}{\sqrt{N}\sigma_1}\right)^N X_k
  \\ \times
  \int_0^\infty dx\, H_{i-1,j}(\nu,x)\,\mathcal{N}(\nu,x)\,f_{lm}(x),
\end{multline}
where
\begin{multline}
  \label{eq:2-59}
  f_{lm}(x) \equiv
  N^N \int d^{N-1}\!W\,\Theta(\lambda_N)\,\lambda_1\cdots\lambda_N\,F_{lm}(J_2,J_3)
  \\ \times
  \exp\left[-\frac{(N-1)(N+2)}{4}J_2\right].
\end{multline}

In the formula of Eq.~(\ref{eq:2-41}), limited number of the
coefficients $G_{ijklm}$ are needed. For $X_k$ of Eq.~(\ref{eq:2-52}),
we need only $X_0 = 1$ and $X_1 = -N/2$. For $H_{i-1,j}(\nu,x)$, we
need only $H_{-1,0}$, $H_{-1,1}$, $H_{-1,3}$, $H_{00}$, $H_{02}$,
$H_{11}$ and $H_{20}$. These functions are straightforwardly evaluated
by Eqs.~(\ref{eq:2-33}) and (\ref{eq:2-54}). For $f_{lm}(x)$, we need
only $f_{00}$, $f_{10}$ and $f_{01}$. The necessary functions
$f_{lm}(x)$ are evaluated for $N=1,2,3$ in the following subsection.

\subsection{\label{subsec:SpecFormulas}
  Specific formulas in one-, two- and
  three-dimensional spaces }

\subsubsection{\label{subsubsec:OneDim}
One-dimensional case  
}

In one-dimensional space, $N=1$, only the terms of
$G_{ijk00} \equiv G_{ijk}$ should be retained. The
Eqs.~(\ref{eq:2-41}) and (\ref{eq:2-58}) in this case reduce to
\begin{multline}
  \label{eq:2-61}
  \bar{n}_\mathrm{pk}(\nu)
  = G_{000}
  +
  \frac{\sigma_0}{6}
  \left[
  G_{300} S^{(0)}
  + 4 \gamma G_{210} S^{(1)}
  + 3 \gamma^2 G_{120} S^{(2)}_2
  \right.
  \\
  \left.
  + \gamma^3 G_{030} S^{(3)}_1
  + 4 G_{101} S^{(1)}
  \right],
\end{multline}
and
\begin{equation}
  \label{eq:2-62}
  G_{ijk}(\nu) = \frac{1}{\sqrt{2\pi}}
  \frac{\sigma_2}{\sigma_1} X_k
  \int_0^\infty dx\, H_{i-1,j}(\nu,x)\,\mathcal{N}(\nu,x) f(x),
\end{equation}
where $X_0 = 1$, $X_1 = -1/2$. Putting $N=1$, $(N-1)J_2=0$ and $l=m=0$
in Eq.~(\ref{eq:2-59}), we have
\begin{equation}
  \label{eq:2-64}
  f(x) \equiv f_{00}(x)  = x,
\end{equation}
for $x\geq 0$. The differential number density
$-d\bar{n}_\mathrm{pk}/d\nu$ can be evaluated by replacing
$H_{i-1,j} \rightarrow H_{ij}$ in Eq.~(\ref{eq:2-62}), and the
resulting expression can be found analytically in this 1D case.
Although we do not reproduce the result here, the analytic expression
is straightforwardly obtained by using a software package such as
\textsl{Mathematica}.

\subsubsection{\label{subsubsec:TwoDim}
Two-dimensional case  
}

In two-dimensional space, $N=2$, only the terms of
$G_{ijkl0} \equiv G_{ijkl}$ should be retained. The
Eqs.~(\ref{eq:2-41}) and (\ref{eq:2-58}) in this case reduce to
\begin{multline}
  \label{eq:2-71}
  \bar{n}_\mathrm{pk}(\nu)
  = G_{0000}
  +
  \frac{\sigma_0}{6}
  \left[
  G_{3000} S^{(0)}
  + 4 \gamma G_{2100} S^{(1)}
  \right.
  \\
  + 3 \gamma^2 G_{1200} S^{(2)}_2
  + \gamma^3 G_{0300} S^{(3)}_1
  + 4 G_{1010} S^{(1)}
  \\
  + 2\gamma G_{0110} S^{(2)}
  + 6 \gamma^2 G_{1001} \left( S^{(2)}_2 - S^{(2)}\right)
  \\
  \left.
  + 6 \gamma^3 G_{0101}
  \left(2S^{(3)}_2 - S^{(3)}_1\right)
  \right],
\end{multline}
and
\begin{multline}
  \label{eq:2-72}
  G_{ijkl}(\nu) = \frac{1}{4\pi}
  \left(\frac{\sigma_2}{\sigma_1}\right)^2 X_k
  \int_0^\infty dx\, H_{i-1,j}(\nu,x)\,\mathcal{N}(\nu,x)\,f_l(x),
\end{multline}
where $X_0=1$, $X_1=-1$. To evaluate Eq.~(\ref{eq:2-59}) in the case
of $N=2$, we introduce a set of variables,
\begin{equation}
  \label{eq:2-73}
  x = \lambda_1 + \lambda_2, \quad 
  y = \frac{\lambda_1 - \lambda_2}{2}.
\end{equation}
and we have $J_1=x$, $J_2 = 4y^2$ and $|\det\zeta|=[x^2-(2y)^2]/4$.
The transformation of the volume element, Eq.~(\ref{eq:2-55}), in the
case of $N=2$ results in $dW \propto y\,dy$ \cite{BE87}. Because of
the ordering $\lambda_1 \geq \lambda_2$, the integration domain is
given by $y>0$, and in order to meet the normalization condition,
Eq.~(\ref{eq:2-57}), we have $dW = 8y\,dy$. Thus we have
\begin{equation}
  \label{eq:2-74}
  f_l(x) \equiv f_{l0}(x) =
  8 \int_0^{x/2} dy\,y\, e^{-4y^2}\left(x^2 - 4y^2\right)
  (-1)^l L_l\left(4y^2\right).
\end{equation}
For the evaluation of Eq.~(\ref{eq:2-71}) we need only
\begin{align}
  \label{eq:2-75a}
  f_0(x)
  &= e^{-x^2} + x^2 - 1,
  \\
  \label{eq:2-75b}
  f_1(x)
  &= \left(1 + x^2\right) e^{-x^2} - 1.
\end{align}
The differential number density $-d\bar{n}_\mathrm{pk}/d\nu$ can be
evaluated by replacing $H_{i-1,j} \rightarrow H_{ij}$ in
Eq.~(\ref{eq:2-72}), and the resulting expression can be found
analytically also in this 2D case. Although the resulting expression
is extremely long and we do not reproduce the result here, the analytic
expression is straightforwardly obtained by using a software package
such as \textsl{Mathematica}.

\subsubsection{\label{subsubsec:ThreeDim}
  Three-dimensional case  
}

In three-dimensional space, $N=3$, the
Eqs.~(\ref{eq:2-41}) and (\ref{eq:2-58}) reduce to
\begin{multline}
  \label{eq:2-81}
  \bar{n}_\mathrm{pk}(\nu)
  = G_{00000}
  +
  \frac{\sigma_0}{6}
  \Biggl[
  G_{30000} S^{(0)}
  + 4 \gamma G_{21000} S^{(1)}
  \\
  + 3 \gamma^2 G_{12000} S^{(2)}_2
  + \gamma^3 G_{03000} S^{(3)}_1
  + 4 G_{10100} S^{(1)}
  \\
  + \frac{8}{3}\gamma G_{01100} S^{(2)}
  + 6 \gamma^2 G_{10010} \left( S^{(2)}_2 - S^{(2)}\right)
  \\
  + 3 \gamma^3 G_{01010}
  \left(3S^{(3)}_2 - S^{(3)}_1\right)
  \\
  + \frac{75}{14}\gamma^3 G_{00001}
  \left(\frac{5}{3} S^{(3)}_1 - 3 S^{(3)}_2 \right)
  \Biggr],
\end{multline}
and
\begin{multline}
  \label{eq:2-82}
  G_{ijklm}(\nu) = \frac{1}{(2\pi)^{3/2}}
  \left(\frac{\sigma_2}{\sqrt{3}\sigma_1}\right)^3 X_k
  \\ \times
  \int_0^\infty dx\, H_{i-1,j}(\nu,x)\,\mathcal{N}(\nu,x)\,f_{lm}(x),
\end{multline}
where $X_0=1$, $X_1=-3/2$. To evaluate Eq.~(\ref{eq:2-59}) in the case
of $N=3$, we introduce a set of variables \cite{BBKS},
\begin{equation}
  \label{eq:2-83}
  x = \lambda_1 + \lambda_2 + \lambda_3, \quad 
  y = \frac{\lambda_1 - \lambda_3}{2}, \quad
  z = \frac{\lambda_1 - 2\lambda_2 + \lambda_3}{2}
\end{equation}
and we have $J_1=x$, $J_2 = 3y^2 + z^2$, $J_3 = z^3 - 9 y^2 z$ and
$|\det\zeta| = (x-2z)[(x+z)^2 - (3y)^2]/27$. The transformation of the
volume element, Eq.~(\ref{eq:2-55}), in the case of $N=3$ results in
$dW \propto y(y^2-z^2)\,dy\,dz$ \cite{BBKS}. Because of the ordering
$\lambda_1 \geq \lambda_2 \geq \lambda_3$, the integration domain is
given by $-y\leq z\leq y$ and in order to meet the normalization
condition, Eq.~(\ref{eq:2-57}), we have
$dW = (2\pi)^{-1/2}3^25^{5/2}y(y^2-z^2)\,dy\,dz$. Thus we have
\begin{multline}
  \label{eq:2-84}
  f_{lm}(x) =
  \frac{3^25^{5/2}}{\sqrt{2\pi}}
  \left(
    \int_0^{x/4} dy\int_{-y}^ydz +
    \int_{x/4}^{x/2} dy\int_{3y-x}^ydz
  \right)
  \\ \times
  e^{-5(3y^2+z^2)/2}
  (x-2z)\left[(x+z)^2 - (3y)^2\right] y\left(y^2-z^2\right)
  \\ \times
  F_{lm}\left(3y^2+z^2, z^3-9y^2z\right).
\end{multline}
For the evaluation of Eq.~(\ref{eq:2-71}) we need only
\begin{align}
  \label{eq:2-85a}
  f_{00}(x)
  &=
  \frac{x}{2}\left(x^2-3\right)
  \left[
    \mathrm{erf}\left(\frac{1}{2}\sqrt{\frac{5}{2}}\,x\right) 
    + \mathrm{erf}\left(\sqrt{\frac{5}{2}}\,x\right) 
  \right]
  \nonumber \\
  & \quad
  + \sqrt{\frac{2}{5\pi}}
  \left[
    \left(\frac{x^2}{2} - \frac{8}{5}\right) e^{-5x^2/2}
    + \left(\frac{31}{4}x^2 + \frac{8}{5}\right) e^{-5x^2/8}
  \right],
  \\
  \label{eq:2-85b}
  f_{10}(x)
  &= 
  -\frac{3x}{2}
  \left[
    \mathrm{erf}\left(\frac{1}{2}\sqrt{\frac{5}{2}}\,x\right) 
    + \mathrm{erf}\left(\sqrt{\frac{5}{2}}\,x\right) 
    \right]
  \nonumber\\
  & \quad - \frac{12}{5}\sqrt{\frac{2}{5\pi}}
  \Biggl[
    e^{-5x^2/2}
    - \left(1+\frac{15x^2}{8}\right)
      \left(1+\frac{15x^2}{16}\right) e^{-5x^2/8}
  \Biggr],
\\
  \label{eq:2-85c}
  f_{01}(x)
  &= 
  -\frac{21}{25}
  \left[
    \mathrm{erf}\left(\frac{1}{2}\sqrt{\frac{5}{2}}\,x\right) 
    + \mathrm{erf}\left(\sqrt{\frac{5}{2}}\,x\right) 
    \right]
    \nonumber\\
  &\quad
  + \frac{27x}{10}\sqrt{\frac{2}{5\pi}}
  \left[
    \frac{2}{15} e^{-5x^2/2} +
    \left(
       \frac{11}{5} + \frac{x^2}{4} + \frac{5x^4}{16}
     \right) e^{-5x^2/8}
  \right].
\end{align}
The differential number density $-d\bar{n}_\mathrm{pk}/d\nu$ can be
evaluated by replacing $H_{i-1,j} \rightarrow H_{ij}$ in
Eq.~(\ref{eq:2-82}).

The generalized version of multivariate Hermite polynomials, $H_{ij}$
with $i\geq -1$, have analytic expressions: the Eq.~(\ref{eq:2-54})
can be analytically integrated. Therefore, the expression of
$G_{ijklm}(\nu)$ of Eq.~(\ref{eq:2-82}) is just a one-dimensional
integration. 

\section{\label{sec:Correlations}
  Correlations of peaks in weakly non-Gaussian fields
}

\subsection{\label{subsec:CorrGeneral}
  A General formula
}

The lowest-order non-Gaussian correction to the power spectrum of
peaks can be calculated by a method of generalized Wiener-Hermite
expansions \cite{Mat95} which is described in
Appendix~\ref{app:Expansion}. The result is given by
Eq.~(\ref{eq:a-22}). Identifying the biased field $\mathcal{F}$ as the
peak number density $n_\mathrm{pk}$. We have
\begin{multline}
  \label{eq:3-1}
  P_\mathrm{pk}(k)
  = [g_1(\bm{k})]^2 P(k)
  \\
  + \frac{1}{2}
  \int \frac{d^N\!p}{(2\pi)^N}
  \left[g_2(\bm{p},\bm{k}-\bm{p})\right]^2
  P(p) P(|\bm{k}-\bm{p}|)
  \\
  + g_1(\bm{k}) \int \frac{d^N\!p}{(2\pi)^N}
  g_2(\bm{p},\bm{k}-\bm{p})
  B(\bm{p},\bm{k}-\bm{p},-\bm{k})
  + \cdots,
\end{multline}
where
$g_n(\bm{k}_1,\ldots,\bm{k}_n) =
\mathcal{G}_n(\bm{k}_1,\ldots,\bm{k}_n)/\mathcal{G}_0$. Specifically
for peaks, from Eqs.~(\ref{eq:b-11a})--(\ref{eq:b-11c}), we have
\begin{align}
  \label{eq:3-2a}
  g_1(\bm{k})
  &= g_{10000} + g_{01000} k^2,
  \\
  \label{eq:3-2b}
  g_2(\bm{k}_1,\bm{k}_2)
  &= g_{20000}
    + g_{11000}\left({k_1}^2+{k_2}^2\right)
    \nonumber\\
  & \quad
    + g_{02000}{k_1}^2{k_2}^2
    - 2g_{00100}\bm{k}_1\cdot\bm{k}_2
    \nonumber\\
  & \quad
    + \frac{2Ng_{00010}}{N-1}
    \left[(\bm{k}_1\cdot\bm{k}_2)^2 - \frac{1}{N}{k_1}^2{k_2}^2\right],
\end{align}
where
\begin{equation}
  \label{eq:3-3}
  g_{ijklm} \equiv
  \frac{G_{ijklm}}{{\sigma_0}^i{\sigma_1}^{2k}{\sigma_2}^{j+2l+3m}G_{00000}}.
\end{equation}
In the case of one-dimension, $N=1$, the last term of
Eq.~(\ref{eq:3-2b}) should be omitted.
The last coefficients $g_{ijklm}$ is calculated by
Eq.~(\ref{eq:2-58}), or we have
\begin{equation}
  \label{eq:3-4}
  g_{ijklm} =
  \frac{X_k\int_0^\infty
    dx\,H_{i-1,j}(\nu,x)\,\mathcal{N}(\nu,x)\,f_{lm}(x)} 
  {{\sigma_0}^i{\sigma_1}^{2k}{\sigma_2}^{j+2l+3m}
    \int_0^\infty dx\,H_{-1,0}(\nu,x)\,\mathcal{N}(\nu,x)\,f_{00}(x)}.
\end{equation}

The power spectrum of peaks is affected by exclusion effects: the
peaks of a smoothed field cannot be too close to each other. Although
the exclusion effects affect the small-scale behavior of the
correlation function of peaks, the power spectrum of peaks on all
scales is largely affected by the effect
\cite{Bal13,Bal16,CPP18,MC19}. Therefore, the predictions of the
perturbative method in this paper are more robust for the correlation
function of peaks on large scales \cite{MC19}. Once the power spectrum
of peaks, Eq.~(\ref{eq:3-1}) is calculated, the correlation function
of peaks is given by
\begin{equation}
  \label{eq:3-5}
    \xi_\mathrm{pk}(r) =
    \int \frac{d^Nk}{(2\pi)^N} e^{i\bm{k}\cdot\bm{r}} P_\mathrm{pk}(k).
\end{equation}

\subsection{\label{subsec:AngleInt}
  Angular integrations
}

For fast and accurate evaluations of Eq.~(\ref{eq:3-1}), one can
analytically perform angular integrations, and the resulting
expression can be evaluated by one-dimensional Fast-Fourier Transforms
(FFT). In the case of three-dimensions, such a technique is developed
in a context of nonlinear perturbation theory
\cite{SVM16,SV16,MFHB16,FBMH17}. We extend the same technique to the
two-dimensional case below.

For this purpose, we rewrite the expression of
Eq.~(\ref{eq:3-1}) as
\begin{multline}
  \label{eq:3-11}
  P_\mathrm{pk}(k)
  = [g_1(\bm{k})]^2 P(k)
  + \frac{1}{2}
  \int_{\bm{k}_1+\bm{k}_2=\bm{k}}
  \left[g_2(\bm{k}_1,\bm{k}_2)\right]^2
  P(k_1) P(k_2)
  \\
  + g_1(\bm{k})
  \int_{\bm{k}_1+\bm{k}_2=\bm{k}}
  g_2(\bm{k}_1,\bm{k}_2)
  B(\bm{k}_1,\bm{k}_2,-\bm{k}_1-\bm{k}_2)
  + \cdots,
\end{multline}
where we use a simplified notation,
\begin{equation}
  \label{eq:3-12}
  \int_{\bm{k}_1+\bm{k}_2=\bm{k}} \cdots
  \equiv
  \int \frac{d^Nk_1}{(2\pi)^N} \frac{d^Nk_2}{(2\pi)^N}
  (2\pi)^N\delta_\mathrm{D}^N(\bm{k}_1 + \bm{k}_2 - \bm{k}) \cdots.
\end{equation}

Because of the rotational symmetry, the integrands in second and third
terms besides the delta function are functions of only $k_1$, $k_2$
and $\hat{\bm{k}}_1\cdot\hat{\bm{k}}_2$, where
$\hat{\bm{k}}_i \equiv \bm{k}_i/|\bm{k}_i|$. The factor
$g_2(\bm{k}_1,\bm{k}_2)$ and its square are given by a superposition
of a form $(\hat{\bm{k}}_1\cdot\hat{\bm{k}}_2)^lX(k_1)Y(k_2)$, where
$l$ is a non-negative integer. When the bispectrum
$B(\bm{k}_1,\bm{k}_2,-\bm{k}_1-\bm{k}_2)$ is also given by a
superposition the same form, the integrals in Eq.~(\ref{eq:3-11}) are
given by a superposition of integrals with the following form:
\begin{multline}
  \label{eq:3-13}
  \int_{\bm{k}_1+\bm{k}_2=\bm{k}}
  (\hat{\bm{k}}_1\cdot\hat{\bm{k}}_2)^l
  X(k_1)Y(k_2)
  \\
  = \int d^N\!r\, e^{-i\bm{k}\cdot\bm{r}}
  \int \frac{d^Nk_1}{(2\pi)^N} \frac{d^Nk_2}{(2\pi)^N}
  e^{i(\bm{k}_1+\bm{k}_2)\cdot\bm{r}}
  (\hat{\bm{k}}_1\cdot\hat{\bm{k}}_2)^l
  X(k_1)Y(k_2).
\end{multline}
The angular integration of the above integral is analytically possible
as follows. First, we notice that the integrals over $\bm{k}_1$ and
$\bm{k}_2$ on the right-hand side give a function of $r$ due to
rotational symmetry. Therefore, one can replace factors
$e^{-i\bm{k}\cdot\bm{r}}$ and $e^{i(\bm{k}_1+\bm{k}_2)\cdot\bm{r}}$ by
their averages over angle of $\bm{r}$. In two- and three-dimensions,
we have
\begin{align}
  \label{eq:3-14a}
 &   e^{-i\bm{k}\cdot\bm{r}} \rightarrow J_0(kr), 
 &&  e^{i(\bm{k}_1+\bm{k}_2)\cdot\bm{r}} \rightarrow
    J_0\left(|\bm{k}_1+\bm{k}_2|r\right),
 &&\mathrm{(2D)}, \\
  \label{eq:3-14b}
 &  e^{-i\bm{k}\cdot\bm{r}} \rightarrow j_0(kr)
 && e^{i(\bm{k}_1+\bm{k}_2)\cdot\bm{r}} \rightarrow
    j_0\left(|\bm{k}_1+\bm{k}_2|r\right),
 &&\mathrm{(3D)},
\end{align}
where $J_n(x)$ and $j_n(x)$ are Bessel functions and spherical Bessel
functions, respectively.

\subsubsection{\label{subsubsec:TwoDimAng}
Two-dimensional case
}

In two dimensions, the integral of Eq.~(\ref{eq:3-13}) reduces to
\begin{multline}
  \label{eq:3-21}
  \int_{\bm{k}_1+\bm{k}_2=\bm{k}}
  (\hat{\bm{k}}_1\cdot\hat{\bm{k}}_2)^l
  X(k_1)Y(k_2)
  = 2\pi \int r\,dr\,J_0(kr)
  \\ \times
  \int \frac{d^2k_1}{(2\pi)^2} \frac{d^2k_2}{(2\pi)^2}
  J_0\left(|\bm{k}_1+\bm{k}_2|r\right)
  (\hat{\bm{k}}_1\cdot\hat{\bm{k}}_2)^l
  X(k_1)Y(k_2).
\end{multline}
We apply an addition theorem of the Bessel function,
\begin{equation}
  \label{eq:3-22}
  J_0\left(|\bm{k}_1 + \bm{k}_2|r\right) =
  \sum_{n=-\infty}^\infty (-1)^n J_n(k_1r) J_n(k_2r) e^{in\theta_{12}},
\end{equation}
where $\theta_{12}$ is the angle between $\bm{k}_1$ and $\bm{k}_2$,
i.e., $\hat{\bm{k}}_1\cdot\hat{\bm{k}}_2 = \cos\theta_{12}$.
The angular dependence can be written as
\begin{equation}
  \label{eq:3-23}
  \left(\hat{\bm{k}}_1\cdot\hat{\bm{k}}_2\right)^l
  = \frac{1}{2^l} e^{-il\theta_{12}} \sum_{m=0}^l
  \begin{pmatrix}
    l \\ m
  \end{pmatrix} e^{2im\theta_{12}}.
\end{equation}

Substituting the above equations into Eq.~(\ref{eq:3-21}), we have
\begin{multline}
  \label{eq:3-24}
  \int_{\bm{k}_1+\bm{k}_2=\bm{k}}
  (\hat{\bm{k}}_1\cdot\hat{\bm{k}}_2)^l
  X(k_1)Y(k_2)
  = 2\pi \int r\,dr\,J_0(kr)
  \\ \times
  \frac{1}{2^l}
  \sum_{m=0}^l (-1)^{l-2m}
  \begin{pmatrix}
    l \\ m
  \end{pmatrix}
  X_{l-2m}(r) Y_{l-2m}(r),
\end{multline}
where
\begin{align}
  \label{eq:3-25a}
  X_n(r) &\equiv \int\frac{k\,dk}{2\pi} J_n(kr) X(k), \\
  \label{eq:3-25b}
  Y_n(r) &\equiv \int\frac{k\,dk}{2\pi} J_n(kr) Y(k).
\end{align}
The last integrals are the one-dimensional Hankel transforms, which
can be efficiently evaluated with the one-dimensional FFT using a
software package \textsl{FFTLog} \cite{Ham00}.

Adopting the formula of Eq.~(\ref{eq:3-24}) in the explicit expression
of Eq.~(\ref{eq:3-11}), the power spectrum of peaks,
$P_\mathrm{pk}(k)$, can be evaluated by using the 1D FFT. The correlation
function of peaks, Eq.~(\ref{eq:3-5}), is also evaluated by
\begin{equation}
  \label{eq:3-26}
    \xi_\mathrm{pk}(r) =
    \int \frac{k\,dk}{2\pi} J_0(kr) P_\mathrm{pk}(k).
\end{equation}

\subsubsection{\label{subsubsec:ThreeDimAng}
Three-dimensional case
}

In three dimensions, the integral of Eq.~(\ref{eq:3-13}) reduces to
\begin{multline}
  \label{eq:3-31}
  \int_{\bm{k}_1+\bm{k}_2=\bm{k}}
  (\hat{\bm{k}}_1\cdot\hat{\bm{k}}_2)^l
  X(k_1)Y(k_2)
  = 4\pi \int r^2dr\,j_0(kr)
  \\ \times
  \int \frac{d^3k_1}{(2\pi)^3} \frac{d^3k_2}{(2\pi)^3}
  j_0\left(|\bm{k}_1+\bm{k}_2|r\right)
  (\hat{\bm{k}}_1\cdot\hat{\bm{k}}_2)^l
  X(k_1)Y(k_2).
\end{multline}
We apply an addition theorem of the Bessel function,
\begin{equation}
  \label{eq:3-32}
  j_0\left(|\bm{k}_1 + \bm{k}_2|r\right) =
  \sum_{n=0}^\infty (-1)^n(2n+1) j_n(k_1r) j_n(k_2r) P_n(\cos\theta_{12}),
\end{equation}
where $\theta_{12}$ is the angle between $\bm{k}_1$ and $\bm{k}_2$,
i.e., $\hat{\bm{k}}_1\cdot\hat{\bm{k}}_2 = \cos\theta_{12}$, and
$P_m(\mu) = (2^m m)^{-1}(d/dx)^m[(x^2-1)^m]$ are Legendre polynomials,
which satisfy the orthogonality relation,
\begin{equation}
  \label{eq:3-33}
  \frac{1}{2}\int_{-1}^1 d\mu P_n(\mu)P_m(\mu) = \frac{\delta_{nm}}{2n+1}.
\end{equation}

The angular dependence can be written as
\begin{equation}
  \label{eq:3-34}
  \left(\hat{\bm{k}}_1\cdot\hat{\bm{k}}_2\right)^l =
  \sum_{m=0}^l (2m+1) \alpha_{lm} P_m(\cos\theta_{12}),
\end{equation}
where
\begin{align}
  \label{eq:3-35}
  \alpha_{lm}
  &\equiv
    \frac{1}{2} \int_{-1}^1 d\mu\,\mu^l P_m(\mu)
    \nonumber\\
  &= 
  \begin{cases}
    \displaystyle
    \frac{l!}{2^{(l-m)/2}[(l-m)/2]!\,(l+m+1)!!}
    & \left(
      \begin{matrix} l\geq m,\\ l+m=\mathrm{even}
        \end{matrix}
        \right), \\
    0 & (\mathrm{otherwise}).
  \end{cases}
\end{align}

Substituting Eqs.~(\ref{eq:3-32}) and (\ref{eq:3-34}) into
Eq.~(\ref{eq:3-31}), we have
\begin{multline}
  \label{eq:3-36}
  \int_{\bm{k}_1+\bm{k}_2=\bm{k}}
  (\hat{\bm{k}}_1\cdot\hat{\bm{k}}_2)^l
  X(k_1)Y(k_2)
  = 4\pi \int r^2dr\,j_0(kr)
  \\ \times
  \sum_{m=0}^l (-1)^m (2m+1) \alpha_{lm}
  X_m(r) Y_m(r),
\end{multline}
where
\begin{align}
  \label{eq:3-37a}
  X_m(r) &\equiv \int\frac{k^2dk}{2\pi^2} j_m(kr) X(k), \\
  \label{eq:3-37b}
  Y_m(r) &\equiv \int\frac{k^2dk}{2\pi^2} j_m(kr) Y(k).
\end{align}
The last integrals are the one-dimensional Hankel transforms, which
can be efficiently evaluated with the one-dimensional FFT.

Adopting the formula of Eq.~(\ref{eq:3-36}) in the explicit expression
of Eq.~(\ref{eq:3-11}), the power spectrum of peaks,
$P_\mathrm{pk}(k)$, can be evaluated by using the 1D FFT. The correlation
function of peaks, Eq.~(\ref{eq:3-5}), is also evaluated by
\begin{equation}
  \label{eq:3-38}
    \xi_\mathrm{pk}(r) =
    \int \frac{k^2dk}{2\pi^2} j_0(kr) P_\mathrm{pk}(k).
\end{equation}

\section{\label{sec:GravInst}%
  Weak non-Gaussianity due to nonlinear evolutions in the large-scale
  structure }

In this section, we numerically calculate the formulas derived in
previous sections when the weak non-Gaussianity is evaluated by
nonlinear perturbation theory of gravitational instability in the
large-scale structure of the Universe. In the numerical evaluations
below, the power spectrum of the three-dimensional density field is
calculated by a Boltzmann code \textsl{CLASS} \cite{class11,CLASS}
with a flat $\Lambda$CDM model and cosmological parameters $h=0.6732$,
$\Omega_{\mathrm{b}0}h^2=0.02238$, $\Omega_{\mathrm{cdm}}h^2=0.1201$,
$n_\mathrm{s}=0.9660$, $\sigma_8=0.8120$ (Planck 2018
\cite{Planck2018}).

\subsection{\label{subsec:ThreeDimAbun}%
  The number density of peaks in a three-dimensional density field
  with weak non-Gaussianity induced by gravity}

In a three-dimensional space, we consider an example of peaks in the
dark matter distribution in three-dimensional space. When the peaks of
matter density field is considered, we first smooth the density field
with a smoothing kernel $W(kR)$ in Fourier space, where $R$ is the
smoothing radius. The field variable $\tilde{f}(\bm{k})$ in Fourier
space corresponds to
\begin{equation}
  \label{eq:4-1}
  \tilde{f}(\bm{k}) = W(kR)\, \delta(\bm{k}),
\end{equation}
where $\delta(\bm{k})$ is the density field in Fourier space. In this
paper, we adopt a Gaussian smoothing kernel, $W(kR) = e^{-k^2R^2/2}$.
Denoting the linear power spectrum by $P_\mathrm{L}(k)$ at an
arbitrary redshift, the power spectrum of the smoothed density field
at the lowest order is given by
\begin{equation}
  \label{eq:4-2}
  P(k) = W^2(kR) P_\mathrm{L}(k).
\end{equation}

Adopting the nonlinear perturbation theory of gravitational
instability \cite{BCGS02}, the bispectrum of smoothed matter density
field at the lowest order is given by
\begin{multline}
  \label{eq:4-3}
  B(\bm{k}_1,\bm{k}_2,\bm{k}_3) =
  W(k_1R) W(k_2R) W(k_3R)
  \\ \times
  \left[
    \frac{10}{7} +
    \left(\frac{k_1}{k_2} + \frac{k_2}{k_1}\right)
    \frac{\bm{k}_1\cdot\bm{k}_2}{k_1k_2} +
    \frac{4}{7}
    \left(\frac{\bm{k}_1\cdot\bm{k}_2}{k_1k_2}\right)^2
  \right]
  \\
  \times P_\mathrm{L}(k_1)P_\mathrm{L}(k_2)
  + \mathrm{cyc.}
\end{multline}
The parameters of Eqs.~(\ref{eq:2-42}) are given by integrations of
the bispectrum in a form,
\begin{equation}
  \label{eq:4-4}
  S^{(n)}_j =
  \frac{{\sigma_0}^{2n-4}}{{\sigma_1}^{2n}}
  \int_{\bm{k}_1+\bm{k}_2+\bm{k}_3=\bm{0}}
  s^{(n)}_j(\bm{k}_1,\bm{k}_2,\bm{k}_3)
  B(\bm{k}_1,\bm{k}_2,\bm{k}_3),
\end{equation}
where
\begin{align}
  \label{eq:4-5}
  &
    s^{(0)} = 1, \quad
    s^{(1)} = \frac{3}{4} {k_3}^2, \quad
    s^{(2)} = -\frac{9}{4} (\bm{k}_1\cdot\bm{k}_2){k_3}^2,
    \nonumber\\
  &
    s^{(2)}_2 = {k_1}^2{k_2}^2, \quad
    s^{(3)}_1 = {k_1}^2 {k_2}^2 {k_3}^2, \quad
  s^{(3)}_2 = (\bm{k}_1\cdot\bm{k}_2)^2 {k_3}^2.
\end{align}

Symmetrizing the arguments of $s^{(n)}_j$, and using only the first
term of Eq.~(\ref{eq:4-3}), Eq.~(\ref{eq:4-4}) reduces to an
expression of three-dimensional integrals,
\begin{multline}
  \label{eq:4-6}
  S^{(n)}_j =
  \frac{{\sigma_0}^{2n-4}}{{\sigma_1}^{2n}}
  \int_0^\infty
  \frac{{k_1}^2dk_1}{2\pi^2} \frac{{k_2}^2dk_2}{2\pi^2}
  \\
  \times
  \int_{-1}^1 \frac{d\mu}{2}
  \tilde{s}^{(n)}_j(k_1,k_2,\mu)
  \tilde{B}(k_1,k_2,\mu),
\end{multline}
where
\begin{multline}
  \label{eq:4-7}
  \tilde{B}(k_1,k_2,\mu) \equiv
  3 W(k_1R) W(k_2R) W\left[({k_1}^2 + {k_2}^2 + 2k_1k_2\mu)^{1/2}R\right]
  \\
  \times
  \left[
    \frac{10}{7} + \left(\frac{k_1}{k_2} + \frac{k_2}{k_1}\right)\mu
    + \frac{4}{7}\mu^2
  \right]
  P_\mathrm{L}(k_1)P_\mathrm{L}(k_2),
\end{multline}
and
\begin{align}
  \label{eq:4-8}
  &
    \tilde{s}^{(0)} = 1, \quad
    \tilde{s}^{(1)} =
    \frac{1}{2}
    \left({k_1}^2 + {k_2}^2 + k_1k_2\mu\right),
    \nonumber\\
  &
    \tilde{s}^{(2)} =
    \frac{3}{2} {k_1}^2{k_2}^2(1-\mu^2),
    \nonumber\\
  &
    \tilde{s}^{(2)}_2 =
    \frac{1}{3}
    \left[
      {k_1}^4 + {k_2}^4 + 3{k_1}^2{k_2}^2 +
    2k_1k_2({k_1}^2+{k_2}^2)\mu
    \right],
    \nonumber\\
  &
    \tilde{s}^{(3)}_1 = {k_1}^2 {k_2}^2
    \left({k_1}^2 + {k_2}^2 + 2 k_1 k_2 \mu\right),
    \nonumber\\
  &
    s^{(3)}_2 =
    \frac{1}{3}
    {k_1}^2{k_2}^2
    \left[
    ({k_1}^2 + {k_2}^2)(2\mu^2+1) + 2k_1k_2\mu(\mu^2+2)
    \right].
\end{align}
The integrals of Eq.~(\ref{eq:4-6}) with Eqs.~(\ref{eq:4-7}) and
(\ref{eq:4-8}) are numerically evaluated. Substituting the results
into Eq.~(\ref{eq:2-81}), the number density of peaks
$\bar{n}_\mathrm{pk}(\nu)$ in three dimensions can be evaluated.

\begin{figure}
  \includegraphics[width=21pc]{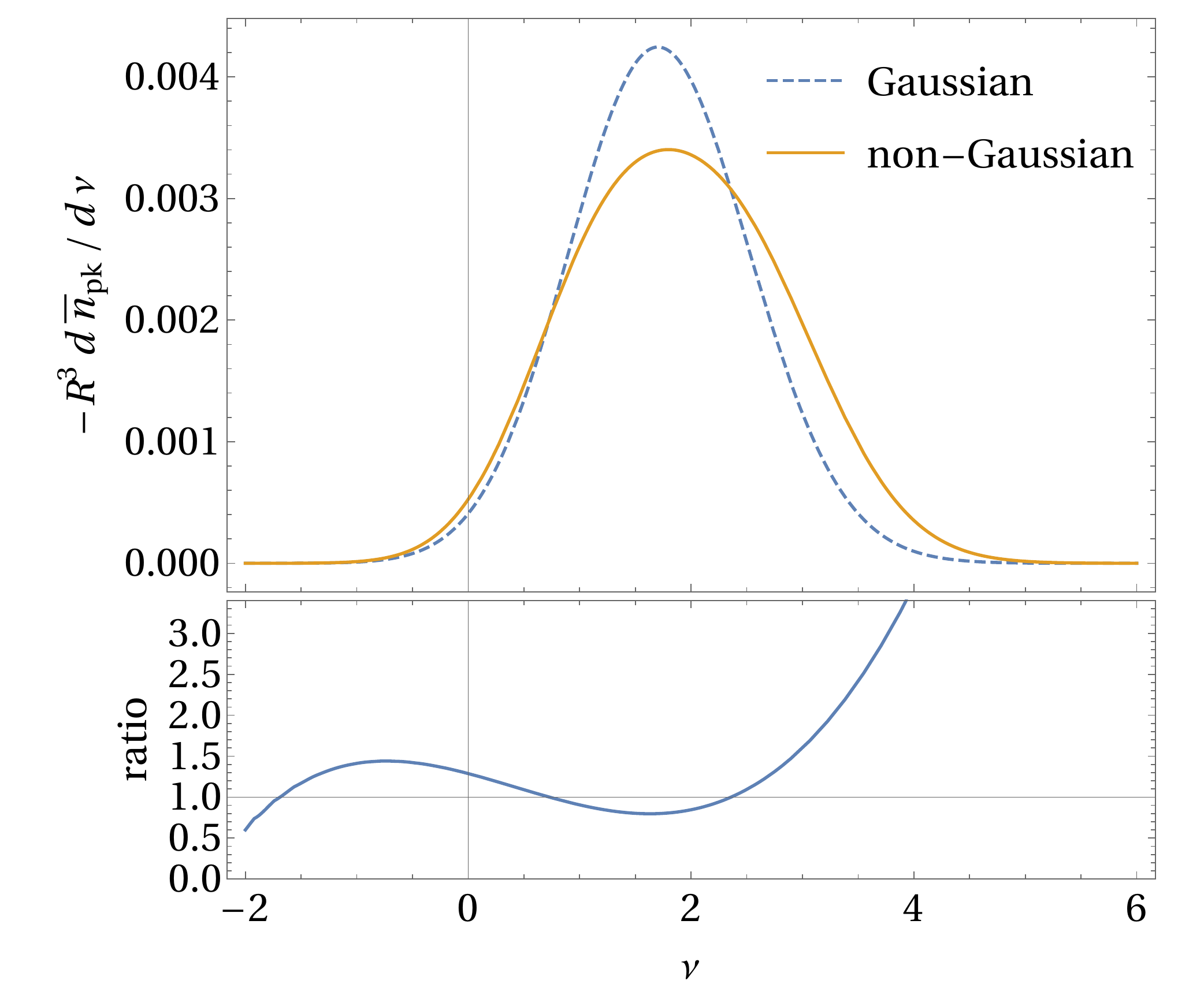}
  \caption{\label{fig:abun3D} The differential number density of peaks
    in three dimensions. In the upper panel, predictions of Gaussian
    (dashed line) and non-Gaussian (solid line) fields are shown,
    where the number density is measured in units of the smoothing
    radius $R = 20\,h^{-1}\mathrm{Mpc}$. In the lower panel, the ratio
    of the non-Gaussian prediction to the Gaussian prediction is
    plotted. }
\end{figure}

In Fig.~\ref{fig:abun3D}, the
differential number density of peaks, $-d\bar{n}_\mathrm{pk}/d\nu$, is
plotted. The Gaussian prediction without the effect of bispectrum is
represented by a dashed line. The gravitational non-Gaussianity
increases the number of high-threshold ($\nu \gtrsim 2.4$) peaks,
because of the positive skewness in the underlying field.

\subsection{\label{subsec:TwoDimAbun}%
  The number density of peaks in a two-dimensional weak lensing field
  with weak non-Gaussianity induced by gravity}

In a two-dimensional space, we consider an example of peaks in the
weak lensing field. When the peaks of weak lensing field is
considered, we first smooth the lensing field with a smoothing kernel
$W(k\vartheta)$, where $\vartheta$ is the smoothing
angle. The field variable $\tilde{f}(\bm{k})$ in Fourier space
corresponds to
\begin{equation}
  \label{eq:4-21}
  \tilde{f}(\bm{k}) = W(k\vartheta) \kappa(\bm{k}),
\end{equation}
where $\kappa(\bm{k})$ is the two-dimensional convergence field of
weak lensing in Fourier space.

For simplicity, we adopt the flat-sky and Limber's approximations
\cite{Lim53} in this paper. Assuming a flat Universe, the power
spectrum and the bispectrum of convergence field are given by
\cite{Kai98,Mat03}
\begin{equation}
  \label{eq:4-22}
  P_\kappa(k) =
  \int \chi^2 d\chi\, q^2(\chi)
  P_\mathrm{3D}\left(\frac{k}{\chi};\chi\right),
\end{equation}
and
\begin{equation}
  \label{eq:4-23}
  B_\kappa\left(k_1,k_2,k_3\right)
  = \int \chi^2 d\chi\, q^3(\chi)
  B_\mathrm{3D}
  \left(
    \frac{k_1}{\chi}, \frac{k_2}{\chi}, \frac{k_3}{\chi}; \chi
  \right),
\end{equation}
where $P_\mathrm{3D}(k,\chi)$ and $B_\mathrm{3D}(k_1,k_2,k_3,\chi)$
are respectively the power spectrum and bispectrum of a
three-dimensional density field at a conformal time, $\tau_0-\chi$
($\tau_0$ is the conformal time at the present), 
\begin{equation}
  \label{eq:4-24}
  q(\chi) \equiv
  \frac{3 {H_0}^2 \Omega_\mathrm{m0}}{2\, a(\chi)}\,
  \frac{\chi_\mathrm{s}-\chi}{\chi\,\chi_\mathrm{s}}
\end{equation}
is a weight function of the convergence field, and $\chi_\mathrm{s}$
is the comoving distance to the source galaxies at a fixed redshift.
In reality, the source redshift has a distribution, and the weight
function should be replaced by an integral over the source redshift.
In this paper we assume a single redshift for source galaxies just for
simplicity.

The two-dimensional power spectrum and bispectrum of the smoothed
convergence field are given by $P(k) = W^2(k\vartheta)P_\kappa(k)$ and
$B(k_1,k_2,k_3) = W(k_1\vartheta) W(k_2\vartheta)
W(k_3\vartheta)B_\kappa(k_1,k_2,k_3)$. The three-dimensional power
spectrum and bispectrum are given by Eqs.~(\ref{eq:4-2}) and
(\ref{eq:4-3}) at the tree level (lowest order) in the perturbation
theory. However, one should apply nonlinear power spectrum and
bispectrum for quantitative predictions for the weak lensing field.
For that purpose, analytic fitting functions of the nonlinear power
spectrum like the Halofit \cite{Smi03,Tak12} and the counterpart of
the nonlinear bispectrum \cite{Laz15,Bos19,Tak19}.

The spectral moments of Eq.~(\ref{eq:2-12}) in the two-dimensional
convergence field are given by
\begin{equation}
  \label{eq:4-25}
  {\sigma_n}^2 =
  \int d\chi\,\chi^{2n+4}q^2(\chi)
  \int \frac{kdk}{2\pi}
  k^{2n} W^2(k\chi\vartheta) P_\mathrm{3D}(k;\chi).
\end{equation}
The skewness parameters of Eq.~(\ref{eq:2-42}) in the two-dimensional
convergence field are given by integrations of the bispectrum in a
form,
\begin{multline}
  \label{eq:4-26}
  S^{(n)}_j =
  \frac{{\sigma_0}^{2n-4}}{{\sigma_1}^{2n}}
  \int d\chi\,\chi^{2n+6} q^3(\chi)
  \int_{\bm{k}_1+\bm{k}_2+\bm{k}_3=\bm{0}}
  s^{(n)}_j(\bm{k}_1,\bm{k}_2,\bm{k}_3)
  \\ \times
  W(k_1\chi\vartheta)W(k_2\chi\vartheta)W(k_3\chi\vartheta)
  B_\mathrm{3D}(\bm{k}_1,\bm{k}_2,\bm{k}_3;\chi),
\end{multline}
where $s^{(n)}_j$ are given by
\begin{align}
  \label{eq:4-27}
  &
    s^{(0)} = 1, \quad
    s^{(1)} = \frac{3}{4} {k_3}^2, \quad
    s^{(2)} = -3 (\bm{k}_1\cdot\bm{k}_2){k_3}^2,
    \nonumber\\
  &
    s^{(2)}_2 = {k_1}^2{k_2}^2, \quad
    s^{(3)}_1 = {k_1}^2 {k_2}^2 {k_3}^2, \quad
  s^{(3)}_2 = (\bm{k}_1\cdot\bm{k}_2)^2 {k_3}^2.
\end{align}
Although we use the same notation $s^{(n)}_j$ as those in
Eq.~(\ref{eq:4-5}) of the three-dimensional case, the coefficient of
$s^{(2)}$ is different in this two-dimensional case, and $\bm{k}_1$,
$\bm{k}_2$, $\bm{k}_3$ are two-dimensional vectors. Integrations over
these vectors are also two-dimensional in Eq.~(\ref{eq:4-26}).

After symmetrizing the arguments of $s^{(n)}_j$, we can replace the
bispectrum $B_\mathrm{3D}$ by an asymmetric counterpart,
$B^\mathrm{asym.}_\mathrm{3D}$, which is defined by
\begin{equation}
  \label{eq:4-28}
  B_\mathrm{3D}(\bm{k}_1,\bm{k}_2,\bm{k}_3;\chi) =
  \frac{1}{3}
  \left[
    B^\mathrm{asym.}_\mathrm{3D}(\bm{k}_1,\bm{k}_2;\chi) +
    \mathrm{cyc.}
  \right].
\end{equation}
Since we have $\bm{k}_1+\bm{k}_2+\bm{k}_3=\bm{0}$, the bispectrum
$B_\mathrm{3D}$ can be always expressible in the form of right-hand
side of Eq.~(\ref{eq:4-28}), even though the choice of functional form
of $B^\mathrm{asym.}_\mathrm{3D}$ is not necessarily unique.

In the case of the tree-level perturbation theory, we have
\begin{equation}
  \label{eq:4-29}
  P_\mathrm{3D}(k;\chi) = D^2(\chi) P_\mathrm{L0}(k)
\end{equation}
and
\begin{multline}
  \label{eq:4-30}
  B^\mathrm{asym.}_\mathrm{3D}(\bm{k}_1,\bm{k}_2;\chi) =
  3D^4(\chi)
  P_\mathrm{L0}(k_1)P_\mathrm{L0}(k_2)
  \\ \times
  \left[
    \frac{10}{7} +
    \left(\frac{k_1}{k_2} + \frac{k_2}{k_1}\right)
    \frac{\bm{k}_1\cdot\bm{k}_2}{k_1k_2} +
    \frac{4}{7}
    \left(\frac{\bm{k}_1\cdot\bm{k}_2}{k_1k_2}\right)^2
  \right],
\end{multline}
where $D(\chi)$ is the linear growth factor at a conformal time
$\tau_0-\chi$ and $P_\mathrm{L0}(k)$ is the linear power spectrum at
the present time. Beyond the tree-level perturbation theory, one can
apply appropriate nonlinear forms of $P_\mathrm{3D}$ and
$B^\mathrm{asym.}_\mathrm{3D}$ instead of Eqs.~(\ref{eq:4-29}) and
(\ref{eq:4-30}), using, e.g., the Halofit approaches. For a
quantitative predictions of the weak lensing field, it is necessary to
adopt nonlinear power spectrum and bispectrum in most of the cases. We
use the tree-level perturbation theory in this paper just for
simplicity.

The skewness parameters of Eq.~(\ref{eq:4-26}) reduces to an
expression,
\begin{multline}
  \label{eq:4-31}
  S^{(n)}_j =
  \frac{{\sigma_0}^{2n-4}}{{\sigma_1}^{2n}}
  \int d\chi\,\chi^{2n+6} q^3(\chi)
  \int \frac{k_1dk_1}{2\pi} \frac{k_2dk_2}{2\pi}
  \\ \times
  \int_{-1}^1 \frac{d\mu}{\pi\sqrt{1-\mu^2}}
  \tilde{s}^{(n)}_j(k_1,k_2,\mu)
  \tilde{B}(k_1,k_2,\mu;\chi),
\end{multline}
where
\begin{multline}
  \label{eq:4-32}
  \tilde{B}(k_1,k_2,\mu;\chi) \equiv
  W(k_1\chi\vartheta)W(k_2\chi\vartheta)
  \\ \times
  W\left[({k_1}^2 + {k_2}^2 + 2k_1k_2\mu)^{1/2}\chi\vartheta\right]
  B^\mathrm{asym.}_\mathrm{3D}(\bm{k}_1,\bm{k}_2;\chi),
\end{multline}
and
\begin{align}
  \label{eq:4-33}
  &
    \tilde{s}^{(0)} = 1, \quad
    \tilde{s}^{(1)} =
    \frac{1}{2}
    \left({k_1}^2 + {k_2}^2 + k_1k_2\mu\right),
    \nonumber\\
  &
    \tilde{s}^{(2)} =
    2 {k_1}^2{k_2}^2(1-\mu^2),
    \nonumber\\
  &
    \tilde{s}^{(2)}_2 =
    \frac{1}{3}
    \left[
      {k_1}^4 + {k_2}^4 + 3{k_1}^2{k_2}^2 +
    2k_1k_2({k_1}^2+{k_2}^2)\mu
    \right],
    \nonumber\\
  &
    \tilde{s}^{(3)}_1 = {k_1}^2 {k_2}^2
    \left({k_1}^2 + {k_2}^2 + 2 k_1 k_2 \mu\right),
    \nonumber\\
  &
    s^{(3)}_2 =
    \frac{1}{3}
    {k_1}^2{k_2}^2
    \left[
    ({k_1}^2 + {k_2}^2)(2\mu^2+1) + 2k_1k_2\mu(\mu^2+2)
    \right].
\end{align}
In the case of the tree-level bispectrum, Eq.~(\ref{eq:4-30}), the
function $\tilde{B}$ of Eq.~(\ref{eq:4-32}) is equivalent to the one
defined in Eq.~(\ref{eq:4-7}) with replacements
$R \rightarrow \chi\vartheta$ and
$P_\mathrm{L}(k) \rightarrow D^2(\chi)P_\mathrm{L0}(k)$. The functions
$\tilde{s}^{n}_j$ in this two-dimensional case are nearly the same as
Eq.~(\ref{eq:4-8}), but the coefficient of $\tilde{s}^{(2)}$ is
different from that in the three-dimensional case.

The integrals of Eq.~(\ref{eq:4-31}) with Eqs.~(\ref{eq:4-32}) and
(\ref{eq:4-33}) are numerically evaluated. For efficient evaluations,
the results of the three-dimensional integrations for fixed values of
$\chi$ are tabulated and interpolated, and finally integrated over
$\chi$. Substituting the results into Eq.~(\ref{eq:2-71}), the number
density of peaks $\bar{n}_\mathrm{pk}(\nu)$ in two dimensions can be
evaluated. In the following example, we simply use the tree-level
power spectrum and bispectrum of Eqs.~(\ref{eq:4-29}) and
(\ref{eq:4-30}) for an illustrative purpose. However, more
quantitative evaluations of the weak lensing field require the use of
nonlinear power spectrum and bispectrum by Halofit etc.

\begin{figure}
  \includegraphics[width=21pc]{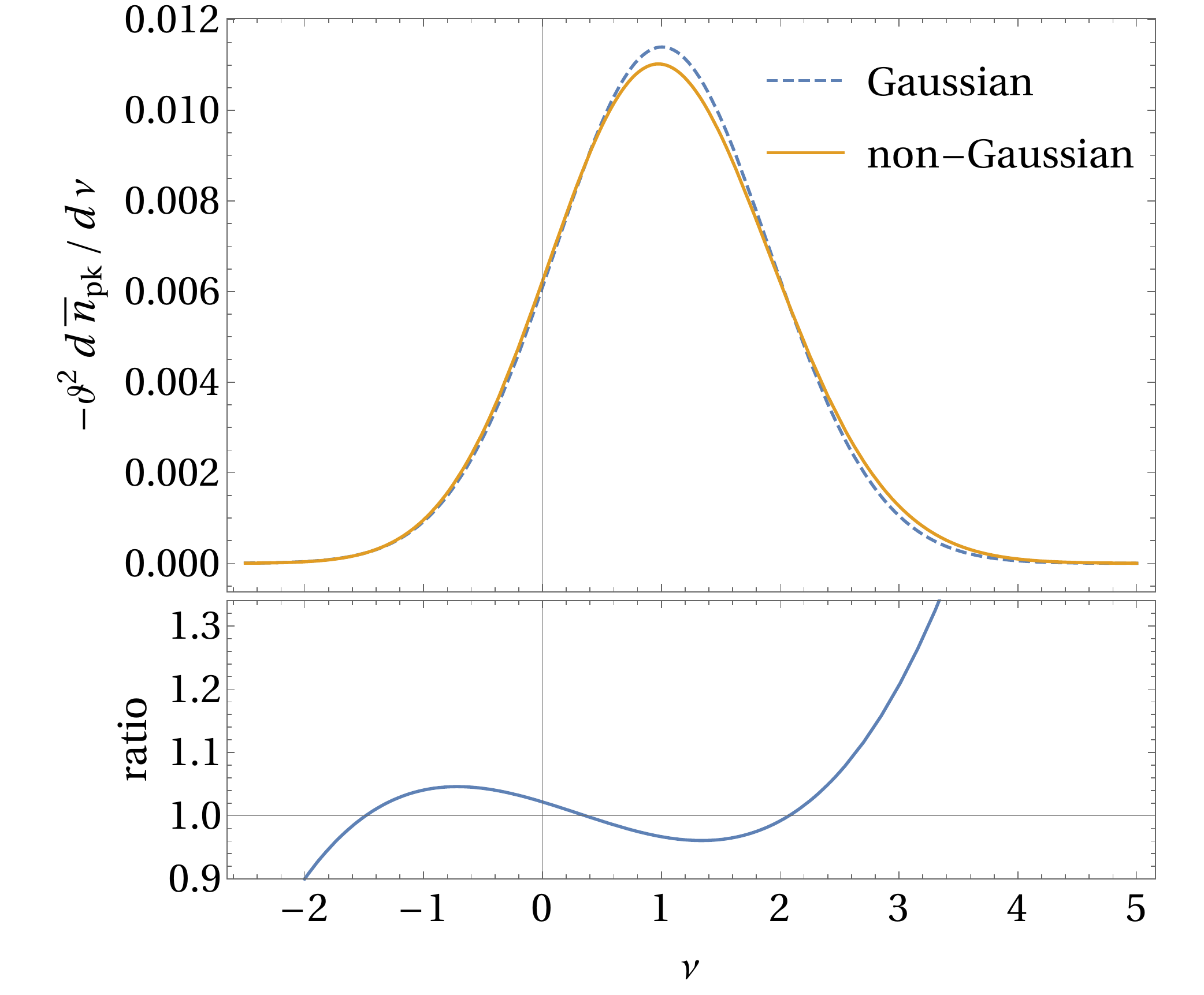}
  \caption{\label{fig:abun2D} The differential number density of peaks
    in two-dimensional weak lensing field. In the upper panel,
    predictions of Gaussian (dashed line) and non-Gaussian (solid
    line) fields are shown, where the number density is measured in
    units of the smoothing angle $\vartheta = 10\,\mathrm{arcmin}$. In
    the lower panel, the ratio of the non-Gaussian prediction to the
    Gaussian prediction is plotted. Nonlinearity and noise effects are
    not included. }
\end{figure}

In Fig.~\ref{fig:abun2D}, the differential number density of peaks in
the weak lensing field, $-d\bar{n}_\mathrm{pk}/d\nu$, is plotted. The
Gaussian prediction without the effect of bispectrum is represented by
a dashed line. We apply the Gaussian smoothing function with a
smoothing angle $\vartheta = 2\,\mathrm{arcmin}$, and the source
redshift is assumed to be fixed at $z_\mathrm{s}=1.5$. In this plot,
we simply use the tree-level predictions of the power spectrum and
bispectrum by the perturbation theory, Eqs.~(\ref{eq:4-29}) and
(\ref{eq:4-30}) as noted above. The gravitational non-Gaussianity
increases the number of high-threshold ($\nu \gtrsim 2$) peaks,
because of the positive skewness in the underlying field.

The shape of the differential number density of peaks relative to the
Gaussian prediction in this plot explains qualitative behavior of the
results from the analysis of numerical simulations presented in
Refs.~\cite{Yan11,Cou19}, although the adopted parameters are
different. In order to quantitatively compare the prediction with the
results of numerical simulations, one needs to use nonlinear fitting
functions for power spectrum and bispectrum, and also needs to take
noise effects into account. It is beyond the scope of this paper to
make detailed comparison with numerical simulations of weak lensing
field, which is one of interesting future applications of this paper.

\subsection{\label{subsec:ThreeDimCorr}
Correlations of peaks with weak non-Gaussianity induced by gravity}

As the last example of numerical demonstration, we consider the
spatial correlation of peaks with weak non-Gaussianity induced by
gravity in three-dimensional space, $N=3$. Substituting
Eqs.~(\ref{eq:4-2}) and (\ref{eq:4-3}) into Eq.~(\ref{eq:3-11}), we
obtain an expression which consists of a superposition of integrals
with a form of Eq.~(\ref{eq:3-36}). Consequently, we need the
functions
\begin{align}
  \label{eq:4-51a}
  \xi^{(n)}_m(r)
  &\equiv
    \int\frac{k^2dk}{2\pi^2} j_m(kr) k^n W^2(kR)P_\mathrm{L}(k),
  \\
  \label{eq:4-51b}
  A^{(n)}_m(r)
  &\equiv
    \int\frac{k^2dk}{2\pi^2} j_m(kr) k^n W(kR)P_\mathrm{L}(k),
  \\
  \label{eq:4-51c}
  B^{(n)}_m(r)
  &\equiv
    \int\frac{k^2dk}{2\pi^2} j_m(kr) k^nW(kR),
\end{align}
to represent the final result. The final expression has the form,
\begin{equation}
  \label{eq:4-52}
  P_\mathrm{pk}(k) =
  4\pi \int r^2dr\,j_0(kr)
  \left[
    \xi_\mathrm{pk}^{(1)}(r) + \xi_\mathrm{pk}^{(2)}(r)
  +
  g_1(k) S_\mathrm{\!NG}(k,r)
  \right],
\end{equation}
where $\xi^{(1)}_\mathrm{pk}(r)$, $\xi^{(2)}_\mathrm{pk}(r)$ and
$S_\mathrm{\!NG}(k,r)$ are polynomials of the functions of
Eqs.~(\ref{eq:4-51a})--(\ref{eq:4-51c}). Their explicit forms are
somehow tedious and given in Appendix~\ref{app:RadialFunc},
Eqs.~(\ref{eq:c-2})--(\ref{eq:c-4}).

\begin{figure}
  \includegraphics[width=20.5pc]{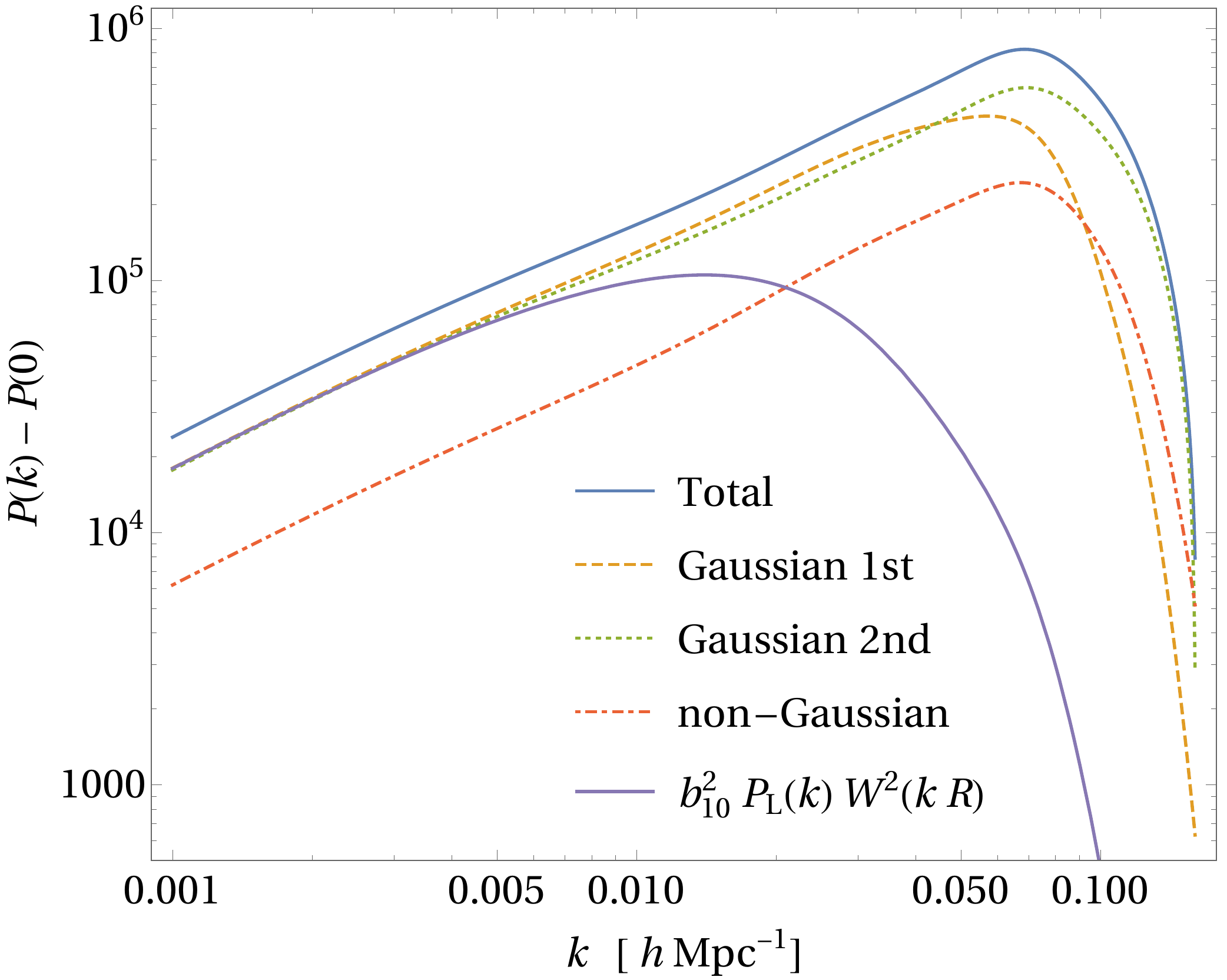}
  \caption{\label{fig:pspec3D} The power spectrum of peaks in
    three-dimensional density field with a smoothing radius
    $R=20\,h^{-1}\mathrm{Mpc}$. Predictions of Gaussian field with
    first-order and second-order approximations are shown in dashed
    and dotted lines, respectively. The component of non-Gaussian
    correction is shown in a dot-dashed line. The total correlation
    function is shown in a solid line. The scaled power spectrum of
    the underlying smoothed density field, ${b_{10}}^2 P_\mathrm{L}(k)
      W^2(kR)$, is also plotted in a lower solid line. }
\end{figure}

For the numerical evaluation of Eq.~(\ref{eq:4-52}), we just need
Hankel transforms, which can be efficiently performed by the use of
\textsl{FFTlog}. In Fig.~\ref{fig:pspec3D}, the result of
Eq.~(\ref{eq:4-52}) is plotted, together with partial components of
the integral. We subtract off the zero-lag value $P(k\rightarrow 0)$
from the power spectrum because of the following reason: As noted in
the last paragraph of Sec.~\ref{subsec:CorrGeneral}, it has been
suggested that the behavior of the correlation function below the
scales of the exclusion zone ($\lesssim R$) non-trivially affects the
power spectrum on large scales ($k\rightarrow 0$)
\cite{Bal13,Bal16,CPP18,MC19}. Accordingly, the second-order
approximation of the power spectrum (the contribution of
$\xi^{(2)}_\mathrm{pk}(r)$ in Eq.~\ref{eq:4-52}) has a non-zero value
in the limit of $k\rightarrow 0$, which corresponds to unphysical
component in the perturbative expansion. To remove this unphysical
effect, we subtract off the zero-lag value $P(k\rightarrow 0)$ from
the second-order approximation of the power spectrum. Other components
do not have the zero-lag value.

The second-order approximation of the power spectrum with Gaussian
components (the first two terms in the integrand of
Eq.~[\ref{eq:4-52}]) is considered to be accurate for $\lesssim 0.1\
h\mathrm{Mpc}^{-1}$ according to the previous analysis \cite{MC19}.
The shape of the non-Gaussian correction is almost proportional to the
Gaussian contribution on most of the scales. Thereby, the total shape
of the peak power spectrum does not change much by the effect of
non-Gaussianity, but the amplitude.

\begin{figure}
  \includegraphics[width=20.5pc]{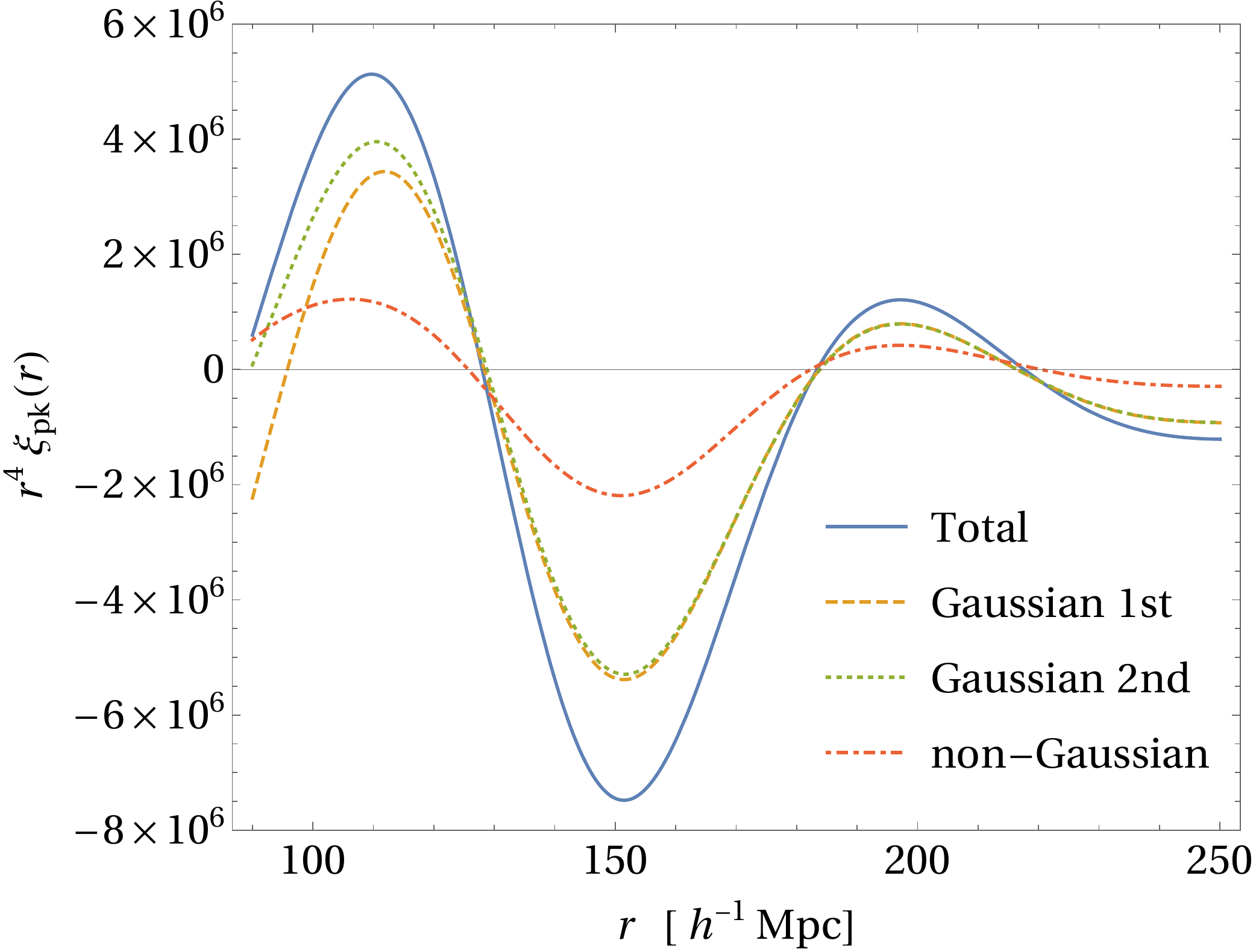}
  \caption{\label{fig:corr3D} The correlation function of peaks in
    three-dimensional density field with a smoothing radius
    $R=20\,h^{-1}\mathrm{Mpc}$. Predictions of Gaussian field with
    first-order and second-order approximations are shown in dotted
    and dashed lines, respectively. The component of non-Gaussian
    correction is shown in dot-dashed line. The total correlation
    function is shown in solid line. }
\end{figure}

\begin{figure}
  \includegraphics[width=20.5pc]{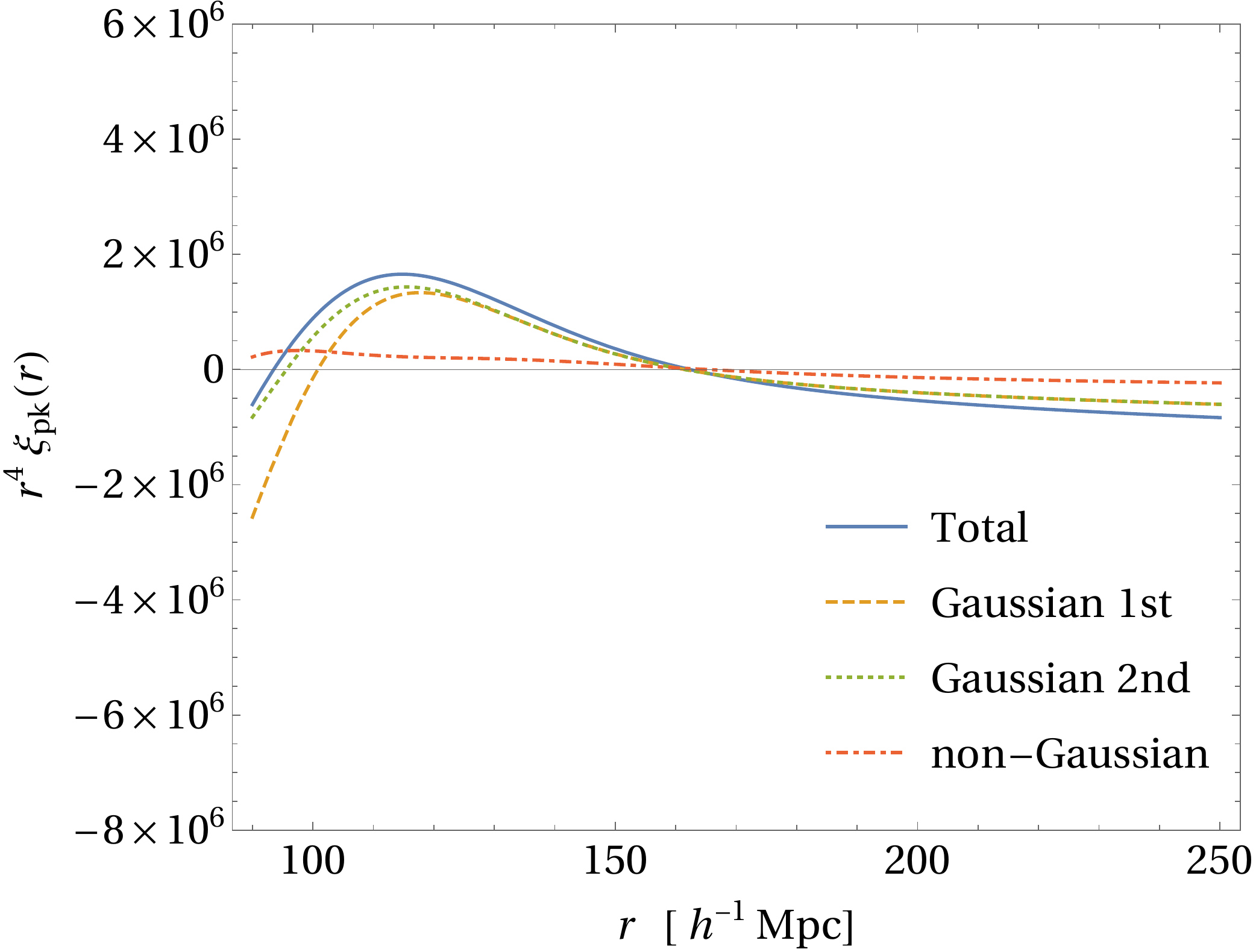}
  \caption{\label{fig:corr3DNB} Same as Fig.~\ref{fig:corr3D} but the
    underlying power spectrum is given by that of CDM power spectrum
    without baryons.}
\end{figure}

Physical implications of the peak clustering are more apparent in
configuration space. The corresponding correlation function,
Eq.~(\ref{eq:3-5}), and its components are plotted in
Fig.~\ref{fig:corr3D}. The vertical axis corresponds to $r^4$ times
the correlation function of peaks. A striking features are the
existence of peaks at around $100\,h^{-1}\mathrm{Mpc}$ and
$200\,h^{-1}\mathrm{Mpc}$, and a trough at around
$150\,h^{-1}\mathrm{Mpc}$. These features are largely due to the
effect of baryon acoustic oscillations (BAO) in the underlying power
spectrum. In fact, if the underlying power spectrum is replaced by the
those of CDM with no baryon, the resulting correlation function is
given by Fig.~\ref{fig:corr3DNB}. The amplitude of the peak around
$100\,h^{-1}\mathrm{Mpc}$ is significantly reduced and the trough and
peak on larger scales both vanish. The fact that baryonic features in
the peak correlation are significantly enhanced is already pointed out
by previous work with Gaussian statistics \cite{Des08,DCSS10}. Here we
see the same property holds with weakly non-Gaussian statistics.


\section{\label{sec:Conclusions}
  Conclusions
}

In this paper, analytic formulas for the statistics of peaks of weakly
non-Gaussian random field are derived. We consider the lowest-order
corrections of non-Gaussianity to the Gaussian predictions, taking the
linear terms of the bispectrum into account. First we generally
consider the statistics of peaks in $N$-dimensional space, and derive
formal expressions of number densities, Eq.~(\ref{eq:2-41}), and the
power spectrum, Eq.~(\ref{eq:3-1}). In order to evaluate the formal
expressions, one need to evaluate $G_{ijklm}$ of Eq.~(\ref{eq:2-58}).
The functions $f_{lm}(x)$ are evaluated in each dimension $N=1,2,3$ as
Eqs.~(\ref{eq:2-64}), (\ref{eq:2-74}) and (\ref{eq:2-84}). The above
equations are our main results of this paper. Useful formulas of
angular integrations to evaluate the power spectrum and the
correlation functions of peaks for $N=2,3$ are given by
Eqs.~(\ref{eq:3-24}) and (\ref{eq:3-36}). In order to illustrate
possible applications of our results, we calculate three examples of
statistics of peaks for cosmological fields: the number density of
peaks in a three-dimensional density field, the number density of
peaks in a two-dimensional weak lensing field, and correlations of
peaks in a three-dimensional density field. In these examples, the
non-Gaussianity is assumed to be induced by nonlinear evolutions of
gravitational instability.

The expansion scheme of the peak abundance by the weak non-Gaussianity
in this paper is equivalent to the pioneering work of
Ref.~\cite{GPP12}. In this previous work, the coefficients of the
expansion for the 3D peaks involves multi-dimensional integrations
which should be evaluated by semi-Monte-Carlo integration. As for the
peak abundance, one of the new developments in this paper is to
provide new formulas for the coefficients, all of which can be
evaluated by virtually one-dimensional integrations. The new formulas
are much easier to evaluate than the previous method, and we believe
they can be widely applied to many problems involving peak statistics
in cosmology. 

As another new development in this paper, we provide new formulas for
the peak correlations in the presence of weak non-Gaussianity. The
methods of deriving general formulas in two and three dimensions are
depicted, and concrete formula in three dimensions with weak
non-Gaussianity induced by gravity is presented [Eqs.~(\ref{eq:4-52})
and (\ref{eq:c-2})--(\ref{eq:c-4})]. Although we do not give the
explicit result, the corresponding formula for 2D lensing field can be
straightforwardly derived.

An interesting feature of the peak correlations of the matter density
field is the enhancement of the effect of BAO in the correlation
function of peaks (Fig.~\ref{fig:corr3D}). Even though the BAO peaks
in the correlation function of the density field is smeared by the
smoothing, the scale of BAO is still encoded in the correlation
function of peaks.

The main purpose of this paper is to provide the analytic formulas for
the peak statistics in the presence of the weak non-Gaussianity. There
are several directions of applying and extending the results of this
paper. First, the peaks of the galaxy number density are obvious sites
of the cosmological structures such as the clusters and superclusters
of galaxies. While the analytic formulas for statistics of peaks in
Gaussian random fields are only applicable in the Lagrangian density
fields, those in weakly non-Gaussian fields are applicable in the
Eulerian density fields which can be directly observable. In the era
of large cosmological surveys, the statistics of peaks in the galaxy
number density fields would be useful tools beyond the two-point
statistics of density fields. Second, analytic formulas of this paper
are also useful for the analysis of 2D weak lensing fields. The weak
lensing fields on scales of interest are definitely non-Gaussian. In
applying the results of this paper, it is necessary to include the
effects of noise, which should be rather straightforward. Third, we
only take into account the effect of lowest-order non-Gaussianity
characterized by the bispectrum. The next-order contributions include
the linear effects of trispectrum and quadratic effects of the
bispectrum. While the next-order contributions of non-Gaussianity are
more complicated than those in this paper, it is feasible to extend
the results in this direction. Fourth, the analysis of the abundance
and correlation of PBH in the presence of initial non-Gaussianity will
be an interesting application of the results of this paper. We hope to
address the possibility of the above applications in near future.

\begin{acknowledgments}
  The author thanks S.~Kuriki for helpful discussion. This work was
  supported by JSPS KAKENHI Grants No.~JP16H03977 and No.~JP19K03835.
\end{acknowledgments}

\newpage
\appendix
\onecolumngrid

\section{\label{app:Expansion} An expansion method with the
  generalized Wiener-Hermite functionals }

In this Appendix, we review a method of Ref.~\cite{Mat95} to derive
the weakly non-Gaussian corrections to the statistical quantities. The
derivation is based on the method of generalized Wiener-Hermite
expansion of the biased field \cite{Mat95}, and this method is closely
related to a method in the integrated perturbation theory
\cite{Mat12}. While the derivation of Ref.~\cite{Mat95} is mostly
presented in configuration space, we present the equivalent method in
Fourier space in this Appendix.

We assume the random field $f$ has a zero-mean,
$\langle f \rangle = 0$, and is statistically homogeneous and
isotropic in $N$-dimensional space. It is convenient to work in
Fourier space and each Fourier coefficients are denoted by
$\tilde{f}(\bm{k})$. Our convention of the Fourier transform is given
by
\begin{equation}
  \label{eq:a-1}
  \tilde{f}(\bm{k}) = \int d^N\!x\,e^{-i\bm{k}\cdot\bm{x}} f(\bm{x}),
  \quad
  f(\bm{x}) = \int \frac{d^N\!k}{(2\pi)^N}\,e^{i\bm{k}\cdot\bm{x}}
  \tilde{f}(\bm{k}).
\end{equation}

The statistical properties are specified by the
probability distribution functional $\mathcal{P}[\tilde{f}]$, which
gives the probability density for a particular form of function
$\tilde{f}(\bm{k})$.

The partition function is given by a functional integral,
\begin{equation}
  \label{eq:a-2}
  \mathcal{Z}[J] =
  \int \mathcal{D}\tilde{f}\,
  \exp\left[i\int \frac{d^N\!k}{(2\pi)^N}
    J(\bm{k}) \tilde{f}(\bm{k})\right]
  \mathcal{P}[\tilde{f}],
\end{equation}
where $\mathcal{D}\tilde{f}$ is the volume element of the functional integral
over the function $\tilde{f}(\bm{k})$ with appropriate measures.
 
According to the cumulant expansion theorem \cite{Ma85}, we have
\begin{equation}
  \label{eq:a-3}
  \ln\mathcal{Z}[J] =
  \sum_{n=1}^{\infty} \frac{i^n}{n!}
  \int \frac{d^N\!k_1}{(2\pi)^N}\cdots \frac{d^N\!k_n}{(2\pi)^N}
  \left\langle
    \tilde{f}(\bm{k}_1)\cdots \tilde{f}(\bm{k}_n)
  \right\rangle_\mathrm{c}
  J(\bm{k}_1) \cdots J(\bm{k}_n),
\end{equation}
where $\langle \cdots \rangle_\mathrm{c}$ represents the $n$-point
cumulant. From the above equation, the partition function is
represented by
\begin{equation}
  \label{eq:a-4}
  \mathcal{Z}[J] =
  \exp\left[
    -\frac{1}{2}\int \frac{d^N\!k}{(2\pi)^N}
     P(k) J(-\bm{k}) J(\bm{k})
  \right]
  \exp\left[
    \sum_{n=3}^\infty \frac{i^n}{n!}
    \int \frac{d^N\!k_1}{(2\pi)^N} \cdots \frac{d^N\!k_n}{(2\pi)^N}
    \left\langle
      \tilde{f}(\bm{k}_1)\cdots \tilde{f}(\bm{k}_n)
    \right\rangle_\mathrm{c}
    J(\bm{k}_1) \cdots J(\bm{k}_n)
  \right],
\end{equation}
where $P(k)$ is the power spectrum defined by
\begin{equation}
  \label{eq:a-5}
  \langle \tilde{f}(\bm{k}_1) \tilde{f}(\bm{k}_2)\rangle_\mathrm{c} =
  (2\pi)^N \delta^N(\bm{k}_1 + \bm{k}_2) P(k).
\end{equation}

Inverting the Eq.~(\ref{eq:a-2}), substituting Eq.~(\ref{eq:a-4}), and
performing Gaussian integration, the probability distribution
functional is represented by
\begin{align}
  \mathcal{P}[\tilde{f}]
  &= \int \bar{\mathcal{D}}J\,
    \mathcal{Z}[J]
    \exp\left[-i\int \frac{d^N\!k}{(2\pi)^N}
    J(\bm{k}) \tilde{f}(\bm{k})\right]
    \nonumber\\
  &=
  \exp\left[
    \sum_{n=3}^\infty \frac{(-1)^n}{n!}
    \int d^N\!k_1 \cdots d^N\!k_N
    \left\langle
    \tilde{f}(\bm{k}_1)\cdots \tilde{f}(\bm{k}_n)
    \right\rangle_\mathrm{c}
    \frac{\delta^n}{\delta \tilde{f}((\bm{k}_1) \cdots \delta
    \tilde{f}(\bm{k}_n)} 
    \right] \mathcal{P}_\mathrm{G}[\tilde{f}],
  \label{eq:a-6}
\end{align}
where $\bar{\mathcal{D}}[J]$ is the volume element of the functional
integral over the function $J(\bm{k})$ with appropriate measures,
$\delta/\delta \tilde{f}(\bm{k})$ is the functional derivative, and
\begin{equation}
  \mathcal{P}_\mathrm{G}[\tilde{f}]
  =
  \int \bar{\mathcal{D}}[J]\,
  \exp\left[
    -\frac{1}{2}\int \frac{d^N\!k}{(2\pi)^N} P(k)
     J(-\bm{k}) J(\bm{k})
    -i\int \frac{d^N\!k}{(2\pi)^N} J(\bm{k}) \tilde{f}(\bm{k})
    \right]
  \propto
    \exp\left[
    -\frac{1}{2}\int
    \frac{d^N\!k}{(2\pi)^N}
    \frac{\tilde{f}(-\bm{k})\tilde{f}(\bm{k})}{P(k)}
    \right]
  \label{eq:a-7}
\end{equation}
is the Gaussian probability distribution functional. The last
expression is the result of the functional integration up to the
normalization constant.

The Eq.~(\ref{eq:a-6}) is a fundamental equation to relate the
non-Gaussian statistics to the Gaussian statistics, and the latter is
analytically easier to calculate than the former in general. In weakly
non-Gaussian cases when the higher-order cumulants are not important,
one can expand the exponent and can investigate the effects of
lower-order cumulants in the non-Gaussian distributions.

Expanding the exponent of Eq.~(\ref{eq:a-6}), we generally have
functional derivatives of $\mathcal{P}_\mathrm{G}$, which is
straightforwardly calculated by the last expression of
Eq.~(\ref{eq:a-7}) and results in polynomials of $\tilde{f}(\bm{k})$
times $\mathcal{P}_\mathrm{G}$. The Wiener-Hermite functionals are the
polynomials of this kind. They are defined by
\begin{equation}
  \label{eq:a-8}
  \mathcal{H}_n(\bm{k}_1,\ldots,\bm{k}_n) \equiv
  \frac{(-1)^n}{\mathcal{P}_\mathrm{G}}
  \frac{\delta^n\mathcal{P}_\mathrm{G}}
  {\delta\tilde{f}(\bm{k}_1)\cdots\delta\tilde{f}(\bm{k}_n)},
\end{equation}
and $\mathcal{H}_0 = 1$ when $n=0$. We also define the dual functionals
$\mathcal{H}^\star_n$ by
\begin{equation}
  \label{eq:a-9}
  \mathcal{H}^\star_n(\bm{k}_1,\ldots,\bm{k}_n) \equiv
  (2\pi)^{Nn}P(k_1) \cdots P(k_n)
  \mathcal{H}_n(-\bm{k}_1,\ldots,-\bm{k}_n).
\end{equation}
The first several functionals are given by
\begin{align}
  \label{eq:a-10a}
  \mathcal{H}^\star_0
  &= 1,\\
  \label{eq:a-10b}
  \mathcal{H}^\star_1(\bm{k})
  &= \tilde{f}(\bm{k}), \\
  \label{eq:a-10c}
  \mathcal{H}^\star_2(\bm{k}_1,\bm{k}_2)
  &= \tilde{f}(\bm{k}_1)\tilde{f}(\bm{k}_2) -
    (2\pi)^N\delta^N(\bm{k}_1+\bm{k}_2) P(k_1),\\
  \label{eq:a-10d}
  \mathcal{H}^\star_3(\bm{k}_1,\bm{k}_2,\bm{k}_3)
  &= \tilde{f}(\bm{k}_1)\tilde{f}(\bm{k}_2)\tilde{f}(\bm{k}_3) -
    \left[
    (2\pi)^N\delta^N(\bm{k}_1+\bm{k}_2) P(k_1)\tilde{f}(\bm{k}_3)
    + \mathrm{cyc.}
    \right],
\end{align}
and so forth. The following orthogonality relation is shown in
Ref.~\cite{Mat95}:
\begin{equation}
  \label{eq:a-11}
  \left\langle
    \mathcal{H}^\star_n(\bm{k}_1,\ldots,\bm{k}_n)
    \mathcal{H}_m(\bm{k}_1',\ldots,\bm{k}_m')
  \right\rangle_\mathrm{G}
  =
  \delta_{nm}
  \left[
    \delta^N(\bm{k}_1-\bm{k}_1')\cdots
    \delta^N(\bm{k}_m-\bm{k}_n') + \mathrm{sym.}(\bm{k}_1,\ldots,\bm{k}_n)
  \right],
\end{equation}
where
$\langle \cdots \rangle_\mathrm{G} = \int
\mathcal{D}\tilde{f}\cdots\mathcal{P}_\mathrm{G}$ is the expectation
value of the Gaussian statistics with the power spectrum $P(k)$, and
$\mathrm{sym.}(\bm{k}_1,\ldots,\bm{k}_n)$ indicates $(n!-1)$ terms to
symmetrize the previous term with respect to permutations of the
arguments $\bm{k}_1,\ldots,\bm{k}_n$. Using the generalized
Wiener-Hermite functionals, the probability distribution functional of
non-Gaussian statistics of Eq.~(\ref{eq:a-6}) is represented by
\begin{multline}
  \label{eq:a-12}
  \mathcal{P} =
  \mathcal{H}_0 \mathcal{P}_\mathrm{G} +
  \frac{1}{6}
  \int d^N\!k_1\,d^N\!k_2\,d^N\!k_3
  \left\langle
    \tilde{f}(\bm{k}_1) \tilde{f}(\bm{k}_2) \tilde{f}(\bm{k}_3)
  \right\rangle_\mathrm{c}
  \mathcal{H}_3(\bm{k}_1,\bm{k}_2,\bm{k}_3)\mathcal{P}_\mathrm{G}
  \\
  + \frac{1}{72}
  \int d^N\!k_1\cdots d^N\!k_6
  \left\langle
    \tilde{f}(\bm{k}_1) \tilde{f}(\bm{k}_2) \tilde{f}(\bm{k}_3)
  \right\rangle_\mathrm{c}
  \left\langle
    \tilde{f}(\bm{k}_4) \tilde{f}(\bm{k}_5) \tilde{f}(\bm{k}_6)
  \right\rangle_\mathrm{c}
  \mathcal{H}_6(\bm{k}_1,\cdots,\bm{k}_6)\mathcal{P}_\mathrm{G}
  \\
  + \frac{1}{24}
  \int d^N\!k_1\cdots d^N\!k_4
  \left\langle
    \tilde{f}(\bm{k}_1) \cdots \tilde{f}(\bm{k}_4)
  \right\rangle_\mathrm{c}
  \mathcal{H}_4(\bm{k}_1,\bm{k}_2,\bm{k}_3,\bm{k}_4)\mathcal{P}_\mathrm{G}
  + \cdots.
\end{multline}
Assuming the higher-order cumulants are small, weakly non-Gaussian
statistics are calculated by the above expansion scheme. This
expansion is a generalization of the Edgeworth expansion
\cite{SB91,Jus95,BK95,Mat95}.

Since the generalized Wiener-Hermite functionals are orthogonal
functionals with orthogonality given by Eq.~(\ref{eq:a-11}), any given
functional $\mathcal{F}(\bm{x})$ of the random field $f$ can be
expanded by the functionals. In Fourier space, the expansion of
$\tilde{\mathcal{F}}(\bm{k})$ is given by
\begin{equation}
  \label{eq:a-13}
  \tilde{\mathcal{F}}(\bm{k}) =
  \sum_{n=0}^\infty \frac{1}{n!}
  \int \frac{d^N\!k_1}{(2\pi)^N}\cdots \frac{d^N\!k_n}{(2\pi)^N}
  (2\pi)^N\delta^N(\bm{k}_1+\cdots +\bm{k}_n-\bm{k})
  \mathcal{G}_n(\bm{k}_1,\ldots,\bm{k}_n)
  \mathcal{H}^\star_n(\bm{k}_1,\ldots,\bm{k}_n),
\end{equation}
where the appearance of the delta function in the integrand is a
consequence of translational invariance of the space. Due to the
orthogonality relation of Eq.~(\ref{eq:a-11}), the coefficient
functions $\mathcal{G}_n$ are given by
\begin{equation}
  \label{eq:a-14}
  (2\pi)^N \delta^N(\bm{k}_1+\cdots +\bm{k}_n - \bm{k})
  \mathcal{G}_{n}\left(\bm{k}_1,\ldots,\bm{k}_n\right) =
  (2\pi)^{Nn}
  \left\langle
    \tilde{\mathcal{F}}(\bm{k}) \mathcal{H}_n(\bm{k}_1,\ldots,\bm{k}_n)
  \right\rangle_\mathrm{G} =
  (2\pi)^{Nn}
  \left\langle
    \frac{\delta^n\tilde{\mathcal{F}}(\bm{k})}
    {\delta \tilde{f}((\bm{k}_1) \cdots \delta \tilde{f}(\bm{k}_n)} 
  \right\rangle_\mathrm{G}.
\end{equation}
Fourier transforming the above equation with respect to $\bm{k}$, we
have
\begin{equation}
  \label{eq:a-15}
  \mathcal{G}_{n}\left(\bm{k}_1,\ldots,\bm{k}_n\right) =
  (2\pi)^{Nn}
  e^{i(\bm{k}_1+\cdot +\bm{k}_n)\cdot\bm{x}}
  \left\langle
    \mathcal{F}(\bm{x}) \mathcal{H}_n(\bm{k}_1,\ldots,\bm{k}_n)
  \right\rangle_\mathrm{G} =
  (2\pi)^{Nn}
  e^{i(\bm{k}_1+\cdot +\bm{k}_n)\cdot\bm{x}}
  \left\langle
    \frac{\delta^n\mathcal{F}(\bm{x})}
    {\delta \tilde{f}((\bm{k}_1) \cdots \delta \tilde{f}(\bm{k}_n)} 
  \right\rangle_\mathrm{G}.
\end{equation}
Due to the translational invariance, the second and third expressions
are independent of the position $\bm{x}$. Thus, in practice, we can
conveniently evaluate the function $\mathcal{G}_n$ by putting
$\bm{x}=\bm{0}$ in the above equation. Using Eqs.~(\ref{eq:a-12}) and
(\ref{eq:a-15}), the expectation value of any functional $\mathcal{F}$
of $f$ at any position is expanded as
\begin{multline}
  \label{eq:a-16}
  \left\langle \mathcal{F} \right\rangle =
  \int \mathcal{D}f\,\mathcal{F}[f]\,\mathcal{P} =
  \mathcal{G}_0 +
  \frac{1}{6}
  \int \frac{d^N\!k_1}{(2\pi)^N} \frac{d^N\!k_2}{(2\pi)^N}
  \frac{d^N\!k_3}{(2\pi)^N}
  \left\langle
    \tilde{f}(\bm{k}_1) \tilde{f}(\bm{k}_2) \tilde{f}(\bm{k}_3)
  \right\rangle_\mathrm{c}
  \mathcal{G}_3(\bm{k}_1,\bm{k}_2,\bm{k}_3)
  \\
  + \frac{1}{24}
  \int \frac{d^N\!k_1}{(2\pi)^N} \cdots \frac{d^N\!k_4}{(2\pi)^N}
  \left\langle
    \tilde{f}(\bm{k}_1) \cdots \tilde{f}(\bm{k}_4)
  \right\rangle_\mathrm{c}
  \mathcal{G}_4(\bm{k}_1,\bm{k}_2,\bm{k}_3,\bm{k}_4)
  \\
  + \frac{1}{72}
  \int \frac{d^N\!k_1}{(2\pi)^N} \cdots \frac{d^N\!k_6}{(2\pi)^N}
  \left\langle
    \tilde{f}(\bm{k}_1) \tilde{f}(\bm{k}_2) \tilde{f}(\bm{k}_3)
  \right\rangle_\mathrm{c}
  \left\langle
    \tilde{f}(\bm{k}_4) \tilde{f}(\bm{k}_5) \tilde{f}(\bm{k}_6)
  \right\rangle_\mathrm{c}
  \mathcal{G}_6(\bm{k}_1,\cdots,\bm{k}_6)
  + \cdots.
\end{multline}
This expansion has a diagrammatic interpretation \cite{Mat95}.
Higher-order correction terms can be efficiently derived by the
diagrammatic rules.

Similarly, we obtain the expansion of the two-point statistics in
Fourier space (the power spectrum),
\begin{equation}
  \label{eq:a-17}
  \left\langle
    \tilde{\mathcal{F}}(\bm{k}) \tilde{\mathcal{F}}(\bm{k}')
  \right\rangle =
  \int \mathcal{D}\tilde{f}\,
  \mathcal{F}(\bm{k})\,\mathcal{F}(\bm{k}')\,\mathcal{P}.
\end{equation}
Substituting Eqs.~(\ref{eq:a-12}) and (\ref{eq:a-13}) into the above
equation, there appear terms involving factors of the type
$\langle \mathcal{H}^\star_n \mathcal{H}^\star_m
\mathcal{H}_l\rangle_\mathrm{G}$. These factors can be evaluated by
applying Wick's theorem for Gaussian statistics, and more conveniently
evaluated by diagrammatic method developed in Ref.~\cite{Mat95}. In
short, contractions of the field which are contained in a same
$\mathcal{H}_n$ or $\mathcal{H}^\star_n$ should be dropped when
applying the Wick's theorem. As a result, the factor
$\langle \mathcal{H}^\star_n \mathcal{H}^\star_m
\mathcal{H}_l\rangle_\mathrm{G}$ is non-zero only when $n+m+l$ is an
even number, and we have, e.g.,
\begin{align}
  \label{eq:a-18a}
  \left\langle
  \mathcal{H}^\star_1(\bm{k}_1)
  \mathcal{H}^\star_1(\bm{k}_1')
  \mathcal{H}_0
  \right\rangle_\mathrm{G}
  &= (2\pi)^NP(k_1)\delta^N(\bm{k}_1+\bm{k}_1'),
  \\
  \label{eq:a-18b}
  \left\langle
  \mathcal{H}^\star_1(\bm{k}_1)
  \mathcal{H}^\star_3(\bm{k}_1',\bm{k}_2',\bm{k}_3')
  \mathcal{H}_0
  \right\rangle_\mathrm{G}
  &= 0,
  \\
  \label{eq:a-18c}
  \left\langle
  \mathcal{H}^\star_2(\bm{k}_1,\bm{k}_2)
  \mathcal{H}^\star_2(\bm{k}_1',\bm{k}_2')
  \mathcal{H}_0
  \right\rangle_\mathrm{G}
  &= (2\pi)^{2N}P(k_1)P(k_2)\delta^N(\bm{k}_1+\bm{k}_1')
    \delta^N(\bm{k}_2+\bm{k}_2') + \mathrm{sym.},
  \\
  \label{eq:a-18d}
  \left\langle
  \mathcal{H}^\star_0
  \mathcal{H}^\star_1(\bm{k}_1)
  \mathcal{H}_3(\bm{k}_1',\bm{k}_2',\bm{k}_3')
  \right\rangle_\mathrm{G}
  &= 0,
  \\
  \label{eq:a-18e}
  \left\langle
  \mathcal{H}^\star_0
  \mathcal{H}^\star_3(\bm{k}_1,\bm{k}_2,\bm{k}_3)
  \mathcal{H}_3(\bm{k}_1',\bm{k}_2',\bm{k}_3')
  \right\rangle_\mathrm{G}
  &=
    \delta^N(\bm{k}_1-\bm{k}_1') \delta^N(\bm{k}_2-\bm{k}_2')
    \delta^N(\bm{k}_3-\bm{k}_3')
    + \mathrm{sym.},
  \\
  \label{eq:a-18f}
  \left\langle
  \mathcal{H}^\star_1(\bm{k}_1)
  \mathcal{H}^\star_2(\bm{k}_1',\bm{k}_2')
  \mathcal{H}_3(\bm{k}_1'',\bm{k}_2'',\bm{k}_3'')
  \right\rangle_\mathrm{G}
  &=
    \delta^N(\bm{k}_1-\bm{k}_1'') \delta^N(\bm{k}_1'-\bm{k}_2'')
    \delta^N(\bm{k}_2'-\bm{k}_3'')
    + \mathrm{sym.},
\end{align}
and so forth, where $+\mathrm{sym.}$ represents the symmetrization
terms which symmetrize the previous term with respect to the arguments
of $\mathcal{H}^\star_n$ or $\mathcal{H}_n$. For example, the
symmetrization terms of Eqs.~(\ref{eq:a-18e}) correspond to cyclic
permutations with respect to $\bm{k}_1$, $\bm{k}_2$ and $\bm{k}_3$.
Substituting Eqs.~(\ref{eq:a-12}) and (\ref{eq:a-13}) into
Eq.~(\ref{eq:a-17}), using Eqs.~(\ref{eq:a-18a})--(\ref{eq:a-18d}),
and subtracting the connected part, we have
\begin{multline}
  \label{eq:a-19}
  \left\langle
    \tilde{\mathcal{F}}(\bm{k}) \tilde{\mathcal{F}}(\bm{k}')
  \right\rangle_\mathrm{c} =
  (2\pi)^N\delta^N(\bm{k}+\bm{k}')
  \mathcal{G}_1(\bm{k}) \mathcal{G}_1(-\bm{k}) P(k)
  \\
  + \frac{1}{2} (2\pi)^N\delta^N(\bm{k}+\bm{k}')
    \int \frac{d^N\!k_1}{(2\pi)^N} \frac{d^N\!k_2}{(2\pi)^N}
    (2\pi)^N\delta^N(\bm{k}_1+\bm{k}_2-\bm{k})
    \mathcal{G}_2(\bm{k}_1,\bm{k}_2)
    \mathcal{G}_2(-\bm{k}_1,-\bm{k}_2)
    P(k_1) P(k_2)
  \\
  + \frac{1}{2}
  \left[
    \mathcal{G}_1(\bm{k})
    \int \frac{d^N\!k_1}{(2\pi)^N} \frac{d^N\!k_2}{(2\pi)^N}
    (2\pi)^N\delta^N(\bm{k}_1+\bm{k}_2-\bm{k}')
    \mathcal{G}_2(\bm{k}_1,\bm{k}_2)
    \left\langle
      \tilde{f}(\bm{k}) \tilde{f}(\bm{k}_1) \tilde{f}(\bm{k}_2)
    \right\rangle_\mathrm{c}
    + (\bm{k} \leftrightarrow \bm{k}')
  \right]
  + \cdots.
\end{multline}
This expression also has a diagrammatic interpretation \cite{Mat95}.
Higher-order correction terms can be efficiently derived by the
diagrammatic rules. In Eqs.~(\ref{eq:a-16}) and (\ref{eq:a-19}), the
$n$-point correlations of the Fourier modes
$\langle\tilde{f}(\bm{k}_1)\cdots\tilde{f}(\bm{k}_n)\rangle_\mathrm{c}$
contains a delta function $\delta^N(\bm{k}_1+\cdots\bm{k}_N)$ due to
the translational invariance of space, and parts of the integrals are
trivially performed. For example, the bispectrum $B$ is defined by
\begin{equation}
  \label{eq:a-20}
  \left\langle
    \tilde{f}(\bm{k}_1) \tilde{f}(\bm{k}_2) \tilde{f}(\bm{k}_3)
  \right\rangle_\mathrm{c}
  = (2\pi)^N\delta^N(\bm{k}_1+\bm{k}_2+\bm{k}_3)
  B(\bm{k}_1,\bm{k}_2,\bm{k}_3).
\end{equation}

The power spectrum $P_\mathcal{F}(k)$ of the biased field
$\mathcal{F}$ is given by
\begin{equation}
  \label{eq:a-21}
  \frac{
    \left\langle
      \tilde{\mathcal{F}}(\bm{k}) \tilde{\mathcal{F}}(\bm{k}')
    \right\rangle_\mathrm{c}
  }{\langle\mathcal{F}\rangle^2}
  = (2\pi)^N\delta^N(\bm{k}+\bm{k}') P_\mathcal{F}(k).
\end{equation}
Thus, from Eqs.~(\ref{eq:a-16}) and (\ref{eq:a-19}) and
(\ref{eq:a-20}), we have
\begin{equation}
  \label{eq:a-22}
  P_\mathcal{F}(k)
  = [g_1(\bm{k})]^2 P(k)
  + \frac{1}{2}
  \int \frac{d^N\!p}{(2\pi)^N}
  \left[g_2(\bm{p},\bm{k}-\bm{p})\right]^2
  P(p) P(|\bm{k}-\bm{p}|)
  + g_1(\bm{k}) \int \frac{d^N\!p}{(2\pi)^N}
  g_2(\bm{p},\bm{k}-\bm{p})
  B(\bm{p},\bm{k}-\bm{p},-\bm{k})
  + \cdots,
\end{equation}
where
\begin{equation}
  \label{eq:a-23}
  g_n(\bm{k}_1,\ldots,\bm{k}_n)
  \equiv \frac{\mathcal{G}_n(\bm{k}_1,\ldots,\bm{k}_n)}{\mathcal{G}_0},
\end{equation}
and we have used a parity symmetry,
$g_n(-\bm{k}_1,\ldots,-\bm{k}_n) = g_n(\bm{k}_1,\ldots,\bm{k}_n)$ and
$B(-\bm{k}_1,-\bm{k}_2,-\bm{k}_3) =B(\bm{k}_1,\bm{k}_2,\bm{k}_3)$. The
non-Gaussian corrections of Eq.~(\ref{eq:a-16}) for the denominator of
Eq.~(\ref{eq:a-21}) do not contribute to the above lowest-order
correction, although they contribute to higher-order corrections in
general. The Eq.~(\ref{eq:a-22}) has the same form as the result of
the integrated perturbation theory \cite{Mat12} if we identify
$g_n = c_n$, where $c_n$ is the renormalized bias function. However,
this identification is valid only for the lowest order non-Gaussian
approximation, because $c_n$ is defined with non-Gaussian statistics
while $g_n$ is defined with Gaussian statistics.

\section{\label{app:Gresp}
  Gaussian response functions for peaks
}

In this Appendix, we calculate the Gaussian $n$-point response
functions of Eq.~(\ref{eq:2-3}) for the peak number density,
$\mathcal{F} = n_\mathrm{pk}$, in a generally $N$-dimensional space. The
functions are defined by
\begin{equation}
  \label{eq:b-1}
  \mathcal{G}_{n}\left(\bm{k}_1,\ldots,\bm{k}_n\right) \equiv
  (2\pi)^{Nn}
  \left\langle
    \frac{\delta^n n_\mathrm{pk}}
    {\delta \tilde{f}(\bm{k}_1) \cdots \delta \tilde{f}(\bm{k}_n)} 
  \right\rangle_\mathrm{G}.
\end{equation}
The peak number density $n_\mathrm{pk}(\nu)$ given by
Eq.~(\ref{eq:2-31}) is a function of the field derivatives, $\alpha$,
$\eta_i$ and $\zeta_{ij}$ of Eq.~(\ref{eq:2-13}). Thereby, the
functional derivative $\delta/\delta\tilde{f}(\bm{k})$ acting on
$n_\mathrm{pk}$ is replaced by
\begin{equation}
  \label{eq:b-2}
  (2\pi)^N \frac{\delta}{\delta\tilde{f}(\bm{k})} \rightarrow
  (2\pi)^N
  \left[
    \frac{\delta\alpha}{\delta\tilde{f}(\bm{k})}
    \frac{\partial}{\partial\alpha} +
    \frac{\delta\eta_i}{\delta\tilde{f}(\bm{k})}
    \frac{\partial}{\partial\eta_i} +
    \frac{\delta\zeta_{ij}}{\delta\tilde{f}(\bm{k})}
    \frac{\partial}{\partial\zeta_{ij}}
    \right] =
  \frac{1}{\sigma_0}\frac{\partial}{\partial\alpha} +
  \frac{ik_i}{\sigma_1}\frac{\partial}{\partial\eta_i} -
  \frac{k_ik_j}{\sigma_2}\frac{\partial}{\partial\zeta_{ij}}
  \equiv \hat{\mathcal{D}}(\bm{k}).
\end{equation}
For the operator $\partial/\partial\zeta_{ij}$, the derivatives are
taken as if $\zeta_{ij}$ and $\zeta_{ji}$ are independent for $i\ne j$
\cite{MD16}. Thus, the Eq.~(\ref{eq:b-1}) is rewritten as
\begin{equation}
  \label{eq:b-3}
  \mathcal{G}_n(\bm{k}_1,\ldots,\bm{k}_n)
  = (-1)^n \int d^{N_Y}\!Y\, n_\mathrm{pk}\,
  \hat{\mathcal{D}}(\bm{k}_1) \cdots \hat{\mathcal{D}}(\bm{k}_n)
  \mathcal{P}_\mathrm{G}(\bm{Y}).
\end{equation}

To calculate the differentiations of the above expression, the
relations
\begin{equation}
  \label{eq:b-4}
  \frac{\partial(\eta^2)}{\partial\eta_i}= 2\eta_i, \quad
  \frac{\partial J_1}{\partial\zeta_{ij}} = -\delta_{ij}, \quad
  \frac{\partial J_2}{\partial\zeta_{ij}} = N\tilde{\zeta}_{ji}, \quad
  \frac{\partial\tilde{\zeta}_{kl}}{\partial\zeta_{ij}} =
  \delta_{ik}\delta_{jl} - \frac{1}{N} \delta_{ij}\delta_{kl}
\end{equation}
are useful. The Gaussian probability distribution function for the
field derivatives $\mathcal{P}_\mathrm{G}(\bm{Y})$ given by
Eq.~(\ref{eq:2-20}) is a function of only rotationally invariant
variables $\alpha$, $\eta^2$, $J_1$ and $J_2$. Using the above
relations, the first-order derivatives are given by
\begin{equation}
  \label{eq:b-5}
  \frac{\partial}{\partial\eta_i} \mathcal{P}_\mathrm{G}
  = 2 \eta_i \frac{\partial}{\partial(\eta^2)} \mathcal{P}_\mathrm{G}, \quad
  \frac{\partial}{\partial\zeta_{ij}} \mathcal{P}_\mathrm{G}
  = \left[
    - \delta_{ij} \frac{\partial}{\partial J_1} 
    + N \tilde{\zeta}_{ji} \frac{\partial}{\partial J_2}
    \right] \mathcal{P}_\mathrm{G},
\end{equation}
the second-order derivatives are given by
\begin{align}
  \label{eq:b-6a}
  \frac{\partial^2}{\partial\eta_i\partial\eta_j} \mathcal{P}_\mathrm{G} &=
  2 \left[
     \delta_{ij} \frac{\partial}{\partial(\eta^2)}
  + 2 \eta_i \eta_j \frac{\partial^2}{\partial(\eta^2)^2}
  \right] \mathcal{P}_\mathrm{G},
\\
  \label{eq:b-6b}
  \frac{\partial^2}{\partial\zeta_{ij}\partial\zeta_{kl}} \mathcal{P}_\mathrm{G}
  &=
  \left[
  \delta_{ij} \delta_{kl} \frac{\partial^2}{\partial {J_1}^2}
  - \frac{2N}{N-1} \left(
    \delta_{ij} \tilde{\zeta}_{lk} + \delta_{kl}
    \tilde{\zeta}_{ji} \right)
  \frac{\partial^2}{\partial J_1 \partial J_2} +
  \frac{4N^2}{(N-1)^2} \tilde{\zeta}_{ji} \tilde{\zeta}_{lk} 
  \frac{\partial^2}{\partial {J_2}^2}
    + \frac{2}{N-1}
    \left(N \delta_{il} \delta_{jk} - \delta_{ij} \delta_{kl}\right)
  \frac{\partial}{\partial J_2}
  \right] \mathcal{P}_\mathrm{G},
\end{align}
and the third-order derivatives are given by
\begin{align}
  \label{eq:b-7a}
  \frac{\partial^3}{\partial\eta_i\partial\eta_j\partial\eta_k} \mathcal{P}_\mathrm{G} &=
  4\left[
    \left(\delta_{ij}\eta_k + \delta_{jk}\eta_i + \delta_{ki}\eta_j\right)
      \frac{\partial^2}{\partial(\eta^2)^2} + 
    2 \eta_i \eta_j \eta_k \frac{\partial^3}{\partial(\eta^2)^3}
  \right] \mathcal{P}_\mathrm{G},
\\
  \label{eq:b-7b}
  \frac{\partial^3}{\partial\zeta_{ij}\partial\zeta_{kl}\partial\zeta_{mn}}
  \mathcal{P}_\mathrm{G} 
  &=
  \left[
  - \delta_{ij} \delta_{kl} \delta_{mn} \frac{\partial^3}{\partial {J_1}^3}
  + \frac{2N}{N-1} \left(
    \delta_{ij} \delta_{kl} \tilde{\zeta}_{nm} +
    \delta_{ij} \delta_{mn} \tilde{\zeta}_{lk} +
    \delta_{kl} \delta_{mn} \tilde{\zeta}_{ji}
    \right)
  \frac{\partial^3}{\partial {J_1}^2 \partial J_2}
  \right.
\nonumber\\
  & \qquad
  -\frac{4N^2}{(N-1)^2} \left(
    \delta_{ij} \tilde{\zeta}_{lk}  \tilde{\zeta}_{nm} +
    \delta_{kl} \tilde{\zeta}_{ji}  \tilde{\zeta}_{nm} +
    \delta_{mn} \tilde{\zeta}_{ji}  \tilde{\zeta}_{lk}
    \right)
  \frac{\partial^3}{\partial J_1 \partial {J_2}^2}
\nonumber\\
  & \qquad
  + \frac{8N^3}{(N-1)^3} \tilde{\zeta}_{ji} \tilde{\zeta}_{lk}
    \tilde{\zeta}_{nm} \frac{\partial^3}{\partial {J_2}^3} +
  \frac{2N}{N-1} \left(
    \frac{3}{N} \delta_{ij} \delta_{kl} \delta_{mn} -
    \delta_{ij} \delta_{kn} \delta_{lm} -
    \delta_{kl} \delta_{in} \delta_{jm} -
    \delta_{il} \delta_{jk} \delta_{mn}
    \right)
  \frac{\partial^2}{\partial J_1 \partial J_2}
\nonumber\\
  & \qquad
  \left. +\,
  \frac{4N^2}{(N-1)^2} \left(
    \delta_{in} \delta_{jm} \tilde{\zeta}_{lk}  +
    \delta_{kn} \delta_{lm} \tilde{\zeta}_{ji}  +
    \delta_{il} \delta_{jk} \tilde{\zeta}_{nm}  -
    \frac{\delta_{ij} \delta_{kl} \tilde{\zeta}_{nm} +
    \delta_{ij} \delta_{mn} \tilde{\zeta}_{lk} +
    \delta_{kl} \delta_{mn} \tilde{\zeta}_{ji}}{N}
    \right)
  \frac{\partial^2}{\partial {J_2}^2}
  \right] \mathcal{P}_\mathrm{G}.
\end{align}
The integrand of Eq.~(\ref{eq:b-3}) other than the product of
operators,
$\hat{\mathcal{D}}(\bm{k}_1)\cdots\hat{\mathcal{D}}(\bm{k}_n)$,
contains only rotationally invariant variables. Thus we can first
average over the angular dependence in the product of operators.
Denoting the angular average by $\langle\cdots\rangle_\Omega$, we have
\begin{align}
  \label{eq:b-8a}
  &
  \left\langle \eta_i \right\rangle_\Omega = 0, \quad
  \left\langle \eta_i \eta_j \right\rangle_\Omega = \frac{1}{N}
  \delta_{ij} \eta^2, \quad
  \left\langle \tilde{\zeta}_{ij} \right\rangle_\Omega = 0, \quad
  \left\langle \tilde{\zeta}_{ij} \tilde{\zeta}_{kl}
  \right\rangle_\Omega =
  \frac{J_2}{N(N+2)}
  \left(
      \delta_{ik} \delta_{jl} +  \delta_{il} \delta_{jk}
      - \frac{2}{N} \delta_{ij} \delta_{kl}
    \right),
\\
  \label{eq:b-8b}
  &
  \left\langle \tilde{\zeta}_{ij} \tilde{\zeta}_{kl} \tilde{\zeta}_{mn}
  \right\rangle_\Omega =
  \frac{J_3}{N^3(N+2)(N+4)}
  \left[
    16 \delta_{ij} \delta_{kl} \delta_{mn} -
    4N \left(
    \delta_{ij} \delta_{km} \delta_{ln} +
    \delta_{ij} \delta_{kn} \delta_{lm} +
    \delta_{kl} \delta_{im} \delta_{jn} +
    \delta_{kl} \delta_{in} \delta_{jm} +
    \delta_{mn} \delta_{ik} \delta_{jl} +
    \delta_{mn} \delta_{il} \delta_{jk}
    \right)
    \right.
\nonumber\\
  & \hspace{7pc}
  \left. +\,
    N^2 \left(
    \delta_{ik} \delta_{lm} \delta_{jn} +
    \delta_{jk} \delta_{lm} \delta_{in} +
    \delta_{il} \delta_{km} \delta_{jn} +
    \delta_{ik} \delta_{ln} \delta_{jm} +
    \delta_{jl} \delta_{km} \delta_{in} +
    \delta_{il} \delta_{kn} \delta_{jm} +
    \delta_{jk} \delta_{ln} \delta_{im} +
    \delta_{jl} \delta_{kn} \delta_{im}
    \right)
    \right],
\end{align}
due to rotaional symmetry. Using the above equations, we have
\begin{align}
  \label{eq:b-9a}
  \left\langle
  \mathcal{D}(\bm{k}) \mathcal{P}_\mathrm{G}
  \right\rangle_\Omega 
  &=
    \left(
    \frac{1}{\sigma_0} \frac{\partial}{\partial\alpha} 
    + \frac{k^2}{\sigma_2} \frac{\partial}{\partial J_1}
    \right) \mathcal{P}_\mathrm{G},
\\
  \label{eq:b-9b}
  \left\langle
  \mathcal{D}(\bm{k}_1) \mathcal{D}(\bm{k}_2)
  \mathcal{P}_\mathrm{G}
  \right\rangle_\Omega
  &=
   \left\{
    \left(
      \frac{1}{\sigma_0} \frac{\partial}{\partial\alpha} 
      + \frac{{k_1}^2}{\sigma_2} \frac{\partial}{\partial J_1}
    \right)
    \left(
      \frac{1}{\sigma_0} \frac{\partial}{\partial\alpha} 
      + \frac{{k_2}^2}{\sigma_2} \frac{\partial}{\partial J_1}
    \right)
   -
    \frac{2(\bm{k}_1\cdot\bm{k}_2)}{{\sigma_1}^2}
    \left[
      1 + \frac{2}{N} \eta^2 \frac{\partial}{\partial (\eta^2)}
    \right] \frac{\partial}{\partial(\eta^2)}
    \right.
\nonumber\\
  & \qquad
    \left.
    + \frac{2N}{(N-1){\sigma_2}^2}
    \left[(\bm{k}_1\cdot\bm{k}_2)^2 - \frac{1}{N}{k_1}^2 {k_2}^2\right]
      \left[
        1 + \frac{4 J_2}{(N-1)(N+2)} \frac{\partial}{\partial J_2}
      \right] \frac{\partial}{\partial J_2 }
    \right\} \mathcal{P}_\mathrm{G},
\\
  \label{eq:b-9c}
  \left\langle
    \mathcal{D}(\bm{k}_1) \mathcal{D}(\bm{k}_2) \mathcal{D}(\bm{k}_3)
  \right\rangle_\Omega \mathcal{P}_\mathrm{G}
  &=
    \left\{
    \left(
      \frac{1}{\sigma_0} \frac{\partial}{\partial\alpha} 
      + \frac{{k_1}^2}{\sigma_2} \frac{\partial}{\partial J_1}
    \right)
    \left(
      \frac{1}{\sigma_0} \frac{\partial}{\partial\alpha} 
      + \frac{{k_2}^2}{\sigma_2} \frac{\partial}{\partial J_1}
    \right)
    \left(
      \frac{1}{\sigma_0} \frac{\partial}{\partial\alpha} 
      + \frac{{k_3}^2}{\sigma_2} \frac{\partial}{\partial J_1}
    \right)
    \right.
\nonumber\\
  & \qquad
   -
    \frac{2(\bm{k}_1\cdot\bm{k}_2)}{{\sigma_1}^2}
    \left(
      \frac{1}{\sigma_0} \frac{\partial}{\partial\alpha} 
      + \frac{{k_3}^2}{\sigma_2} \frac{\partial}{\partial J_1}
    \right)
    \left[
      1 + \frac{2}{N} \eta^2 \frac{\partial}{\partial (\eta^2)}
    \right] \frac{\partial}{\partial(\eta^2)}
    + \mathrm{cyc.}
\nonumber\\
  & \qquad
    + \frac{2N}{(N-1){\sigma_2}^2}
    \left[(\bm{k}_1\cdot\bm{k}_2)^2 - \frac{1}{N}{k_1}^2 {k_2}^2\right]
    \left(
      \frac{1}{\sigma_0} \frac{\partial}{\partial\alpha} 
      + \frac{{k_3}^2}{\sigma_2} \frac{\partial}{\partial J_1}
    \right)
      \left[
        1 + \frac{4 J_2}{(N-1)(N+2)} \frac{\partial}{\partial J_2}
      \right] \frac{\partial}{\partial J_2}
    + \mathrm{cyc.}
\nonumber\\
  & \qquad
    + \frac{64 N^2}{(N-1)^3(N+2)(N+4){\sigma_2}^3}
    \Biggl[
    (\bm{k}_1\cdot\bm{k}_2)(\bm{k}_2\cdot\bm{k}_3)(\bm{k}_3\cdot\bm{k}_1)
    \nonumber\\
  & \hspace{13pc}
    \left.
    - \frac{ {k_1}^2(\bm{k}_2\cdot\bm{k}_3) + \mathrm{cyc.}}{N}
    + \frac{2}{N^2} {k_1}^2 {k_2}^2 {k_3}^2
    \Biggr]
    J_3 \frac{\partial^3}{{\partial J_2}^3}
    \right\} \mathcal{P}_\mathrm{G}.
\end{align}
According to the form of $\mathcal{P}_\mathrm{G}$ in
Eq.~(\ref{eq:2-20}), we have
\begin{align}
  \label{eq:b-10a}
    \left[
    1 + \frac{2}{N} \eta^2 \frac{\partial}{\partial (\eta^2)}
    \right] \frac{\partial}{\partial(\eta^2)} \mathcal{P}_\mathrm{G}
    &= \frac{N}{2}\left(\eta^2 - 1\right) \mathcal{P}_\mathrm{G}
    = - L^{(N/2-1)}_1\left(\frac{N}{2}\eta^2\right) \mathcal{P}_\mathrm{G},
\\
  \label{eq:b-10b}
    \left[
      1 + \frac{4 J_2}{(N-1)(N+2)} \frac{\partial}{\partial J_2}
    \right] \frac{\partial}{\partial J_2} \mathcal{P}_\mathrm{G}
    &= \frac{(N-1)(N+2)}{4}\left(J_2 - 1\right) \mathcal{P}_\mathrm{G}
    = F_{10}(J_2,J_3) \mathcal{P}_\mathrm{G}, 
\\
  \label{eq:b-10c}
    J_3 \frac{\partial^3}{{\partial J_2}^3} \mathcal{P}_\mathrm{G}
    &= -\frac{(N-1)^3(N+2)^3}{64} J_3 \mathcal{P}_\mathrm{G}
    = -\frac{(N-1)^3(N+2)^3}{64} F_{01}(J_2,J_3) \mathcal{P}_\mathrm{G},
\end{align}
where $F_{lm}(J_2,J_3)$ is defined by Eq.~(\ref{eq:2-35}).
Substituting Eqs.~(\ref{eq:b-10a})--(\ref{eq:b-10c}) into
Eqs.~(\ref{eq:b-9a})--(\ref{eq:b-9c}), the Eq.~(\ref{eq:b-3}) is
represented in terms of $G_{ijklm}$ of Eq.~(\ref{eq:2-32}). The
results are given by
\begin{align}
  \label{eq:b-11a}
  \mathcal{G}_0
  &= G_{00000},
  \\
  \label{eq:b-11b}
  \mathcal{G}_1(\bm{k})
  &= \frac{G_{10000}}{\sigma_0} + \frac{G_{01000}}{\sigma_2}k^2,
  \\
  \label{eq:b-11c}
  \mathcal{G}_2(\bm{k}_1,\bm{k}_2)
  &= \frac{G_{20000}}{{\sigma_0}^2}
    + \frac{G_{11000}}{\sigma_0\sigma_2}\left({k_1}^2+{k_2}^2\right)
    + \frac{G_{02000}}{{\sigma_2}^2}{k_1}^2{k_2}^2
    - \frac{2G_{00100}}{{\sigma_1}^2}\bm{k}_1\cdot\bm{k}_2
    + \frac{2NG_{00010}}{(N-1){\sigma_2}^2}
    \left[(\bm{k}_1\cdot\bm{k}_2)^2 - \frac{1}{N}{k_1}^2{k_2}^2\right],
  \\
  \nonumber
  \mathcal{G}_3(\bm{k}_1,\bm{k}_2,\bm{k}_3)
  &= \frac{G_{30000}}{{\sigma_0}^3}
    + \frac{G_{21000}}{{\sigma_0}^2\sigma_2}
    \left({k_1}^2+{k_2}^2+{k_3}^2\right)
    + \frac{G_{12000}}{\sigma_0{\sigma_2}^2}
    \left({k_1}^2{k_2}^2+\mathrm{cyc.}\right)
  \\
  \nonumber
  & \quad
    + \frac{G_{03000}}{{\sigma_2}^3} {k_1}^2 {k_2}^2 {k_3}^2
    - \frac{2G_{10100}}{\sigma_0{\sigma_1}^2}
    \left(\bm{k}_1\cdot\bm{k}_2 + \mathrm{cyc.}\right)
    - \frac{2G_{01100}}{{\sigma_1}^2\sigma_2}
    \left[(\bm{k}_1\cdot\bm{k}_2){k_3}^2 + \mathrm{cyc.}\right]
  \\
  \nonumber
  & \quad
    + \frac{2NG_{10010}}{(N-1)\sigma_0{\sigma_2}^2}
    \left[(\bm{k}_1\cdot\bm{k}_2)^2 - \frac{1}{N}{k_1}^2{k_2}^2\right]
    + \frac{2NG_{01010}}{(N-1){\sigma_2}^3}
    \left[
    (\bm{k}_1\cdot\bm{k}_2)^2{k_3}^2 + \mathrm{cyc.}
    - \frac{3}{N}{k_1}^2{k_2}^2{k_3}^3
    \right]
  \\
  \label{eq:b-11d}
  & \quad
    - \frac{N^2 (N+2)^2 G_{00001}}{(N+4){\sigma_2}^3}
    \left[
    (\bm{k}_1\cdot\bm{k}_2)(\bm{k}_2\cdot\bm{k}_3)(\bm{k}_3\cdot\bm{k}_1)
    - \frac{(\bm{k}_1\cdot\bm{k}_2)^2{k_3}^2 + \mathrm{cyc.}}{N}
    + \frac{2}{N^2}{k_1}^2{k_2}^2{k_3}^2
    \right].
\end{align}

\section{\label{app:RadialFunc}
  Radial functions for correlations of peaks
}

In this Appendix, radial functions $\xi^{(1)}_\mathrm{pk}(r)$,
$\xi^{(2)}_\mathrm{pk}(r)$ and $S_\mathrm{\!NG}(k,r)$ in
Eq.~(\ref{eq:4-52}) are explicitly given by the functions
$\xi^{(n)}_m(r)$, $A^{(n)}_m(r)$, $B^{(n)}_m(r)$ of
Eqs.~(\ref{eq:4-51a})--(\ref{eq:4-51c}). The relations are derived from
Eqs.~(\ref{eq:3-11}), (\ref{eq:3-36}), (\ref{eq:4-2}) and
(\ref{eq:4-3}). The results for $\xi^{(1)}_\mathrm{pk}(r)$ and
$\xi^{(2)}_\mathrm{pk}(r)$ are already given in Ref.~\cite{MC19}. We
reproduce the latter results here for completeness. The result for
$S_\mathrm{\!NG}(k,r)$ below is new in this paper.

In the following, we adopt notations,
\begin{equation}
  \label{eq:c-1}
  b_{ij} \equiv g_{ij000}, \quad
  \chi_k \equiv g_{00k00}, \quad
  \omega_{lm} \equiv g_{000lm}.
\end{equation}
The final results are given by
\begin{equation}
  \label{eq:c-2}
  \xi_\mathrm{pk}^{(1)}
  = {b_{10}}^2 \xi^{(0)}_0
  + 2 b_{10}b_{01} \xi^{(2)}_0
  + {b_{01}}^2 \xi^{(4)}_0,
\end{equation}
\begin{multline}
  \label{eq:c-3}
  \xi_\mathrm{pk}^{(2)}
  = {b_{20}}^2 \left(\xi^{(0)}_0\right)^2
    + 4 b_{20}b_{11} \xi^{(0)}_0 \xi^{(2)}_0
    + 2 {b_{11}}^2 \xi^{(0)}_0 \xi^{(4)}_0
    + 2 \left(b_{20}b_{02} + {b_{11}}^2 + \frac{2}{3} {\chi_1}^2\right)
    \left(\xi^{(2)}_0\right)^2
    + 4 b_{11}b_{02} \xi^{(2)}_0 \xi^{(4)}_0
    + 4 b_{20} \chi_1 \left(\xi^{(1)}_1\right)^2
    \\
    + 8 b_{11} \chi_1 \xi^{(1)}_1 \xi^{(3)}_1
    + \left( {b_{02}}^2 + \frac{4}{5} {\omega_{10}}^2 \right)
      \left(\xi^{(4)}_0\right)^2
    + 4 \left(b_{02} + \frac{4}{5}\omega_{10}\right)\chi_1
      \left(\xi^{(3)}_1\right)^2
    + 4 \left(b_{20}\omega_{10} + \frac{2}{3} {\chi_1}^2 \right)
      \left(\xi^{(2)}_2\right)^2
    + 8 b_{11} \omega_{10} \xi^{(2)}_2 \xi^{(4)}_2
    \\
    + 4 \left(b_{02} + \frac{2}{7} \omega_{10} \right) \omega_{10}
      \left(\xi^{(4)}_2\right)^2
    + \frac{24}{5} \chi_1 \omega_{10} \left(\xi^{(3)}_3\right)^2
    + \frac{72}{35} {\omega_{10}}^2 \left(\xi^{(4)}_4\right)^2,
\end{multline}
and 
\begin{align}
  S_\mathrm{\!NG} =
  &\,\,
    2 b_{20}
    \left[\frac{17}{21} \left(A^{(0)}_0\right)^2
    + \frac{4}{21} \left(A^{(0)}_2\right)^2
    - A^{(-1)}_1 A^{(1)}_1
    + \frac{3}{7} A^{(0)}_0 B^{(0)}_0
    + A^{(-1)}_1 B^{(1)}_1
    - \frac{1}{7 k^2} A^{(1)}_1 B^{(1)}_1
    + \frac{4}{21 k^2} A^{(0)}_0 B^{(2)}_0
    + \frac{8}{21 k^2} A^{(0)}_2 B^{(2)}_2
  \right]
  \nonumber \\
  &  + 2 b_{11}
  \left[
    \frac{34}{21} A^{(2)}_0 A^{(0)}_0
    - \left(A^{(1)}_1\right)^2
    + \frac{8}{21} A^{(0)}_2 A^{(2)}_2
    - A^{(-1)}_1 A^{(3)}_1
    + \frac{3}{7} A^{(2)}_0 B^{(0)}_0
    + A^{(1)}_1 B^{(1)}_1
    + \frac{3}{7} A^{(0)}_0 B^{(2)}_0
    + A^{(-1)}_1 B^{(3)}_1
  \right.
  \nonumber \\
  &  \left. \hspace{3pc}
    - \frac{3}{7 k^2} A^{(4)}_0 B^{(0)}_0
    - \frac{1}{7 k^2} A^{(3)}_1 B^{(1)}_1
    - \frac{5}{21 k^2} A^{(2)}_0 B^{(2)}_0
    + \frac{8}{21 k^2} A^{(2)}_2 B^{(2)}_2
    - \frac{1}{7 k^2} A^{(1)}_1 B^{(3)}_1
    + \frac{4}{21 k^2} A^{(0)}_0 B^{(4)}_0
    + \frac{8}{21 k^2} A^{(0)}_2 B^{(4)}_2
  \right]
  \nonumber \\
  & + 2 b_{02}
  \left[
    \frac{17}{21} \left(A^{(2)}_0\right)^2
    + \frac{4}{21} \left(A^{(2)}_2\right)^2
    - A^{(1)}_1 A^{(3)}_1
    + \frac{3}{7} A^{(2)}_0 B^{(2)}_0
    + A^{(1)}_1 B^{(3)}_1
  \right.
  \nonumber \\
  & \left. \hspace{14pc}
    - \frac{3}{7 k^2} A^{(4)}_0 B^{(2)}_0
    - \frac{1}{7 k^2} A^{(3)}_1 B^{(3)}_1
    + \frac{4}{21 k^2} A^{(2)}_0 B^{(4)}_0
    + \frac{8}{21 k^2} A^{(2)}_2 B^{(4)}_2
  \right]
  \nonumber \\
  &  + 4 \chi_1
  \left[
    - \frac{1}{3} A^{(2)}_0 A^{(0)}_0
    + \frac{31}{35} \left(A^{(1)}_1\right)^2
    + \frac{4}{35} \left(A^{(1)}_3\right)^2
    - \frac{2}{3} A^{(0)}_2 A^{(2)}_2
    + \frac{3}{7} A^{(1)}_1 B^{(1)}_1
    + \frac{1}{3} A^{(0)}_0 B^{(2)}_0
    + \frac{2}{3} A^{(0)}_2 B^{(2)}_2
  \right.
  \nonumber \\
  &
    \left. \hspace{10pc}
    - \frac{3}{7 k^2} A^{(3)}_1 B^{(1)}_1
    - \frac{1}{21 k^2} A^{(2)}_0 B^{(2)}_0
    - \frac{2}{21 k^2} A^{(2)}_2 B^{(2)}_2
    + \frac{12}{35 k^2} A^{(1)}_1 B^{(3)}_1
    + \frac{8}{35 k^2} A^{(1)}_3 B^{(3)}_3
  \right]
  \nonumber \\
  & + 4 \omega_{10}
  \left[
    \frac{4}{105} \left(A^{(2)}_0\right)^2
    + \frac{127}{147} \left(A^{(2)}_2\right)^2
    + \frac{24}{245} \left(A^{(2)}_4\right)^2
    - \frac{2}{5} A^{(1)}_1 A^{(3)}_1
    - \frac{3}{5} A^{(1)}_3 A^{(3)}_3
    + \frac{3}{7} A^{(2)}_2 B^{(2)}_2
    + \frac{2}{5} A^{(1)}_1 B^{(3)}_1
    + \frac{3}{5} A^{(1)}_3 B^{(3)}_3
    \right.
   \nonumber \\
  &  \left. \hspace{5pc}
    - \frac{3}{7 k^2} A^{(4)}_2 B^{(2)}_2
    - \frac{2}{35 k^2} A^{(3)}_1 B^{(3)}_1
    - \frac{3}{35 k^2} A^{(3)}_3 B^{(3)}_3
    + \frac{8}{105 k^2} A^{(2)}_0 B^{(4)}_0
    + \frac{44}{147 k^2} A^{(2)}_2 B^{(4)}_2
    + \frac{48}{245 k^2} A^{(2)}_4 B^{(4)}_4
    \right].
    \label{eq:c-4}
\end{align}


\renewcommand{\apj}{Astrophys.~J. }
\newcommand{\aap}{Astron.~Astrophys. }
\newcommand{\aj}{Astron.~J. }
\newcommand{\apjl}{Astrophys.~J.~Lett. }
\newcommand{\apjs}{Astrophys.~J.~Suppl.~Ser. }
\newcommand{\apss}{Astrophys.~Space Sci. }
\newcommand{\cqg}{Class.~Quant.~Grav. }
\newcommand{\jcap}{J.~Cosmol.~Astropart.~Phys. }
\newcommand{\mnras}{Mon.~Not.~R.~Astron.~Soc. }
\newcommand{\mpla}{Mod.~Phys.~Lett.~A }
\newcommand{\pasj}{Publ.~Astron.~Soc.~Japan }
\newcommand{\physrep}{Phys.~Rep. }
\newcommand{\ptp}{Progr.~Theor.~Phys. }
\newcommand{\ptep}{Prog.~Theor.~Exp.~Phys. }
\newcommand{\jetp}{JETP }


\begin{thebibliography}{10}
  
\bibitem{Kai84} N.~Kaiser, \apjl \textbf{284}, L9 (1984).
  
\bibitem{DEFW85} M.~Davis, G.~Efstathiou C.~S.~Frenk and
  S.~D.~M.~White, \apj \textbf{292}, 371 (1985).

\bibitem{DR87} A.~Dekel and M.~J.~Rees, Nature \textbf{326}, 455
  (1987).

\bibitem{Dor70} A.~G.~Doroshkevich, Astrofiz. \textbf{6}, 581 (1970).
  
\bibitem{BBKS} J.~M.~Bardeen, J.~R.~Bond, N.~Kaiser and A.~S.~Szalay,
  \apj \textbf{304}, 15 (1986).

\bibitem{PH85} J.~A.~Peacock and A.~F.~Heavens, \mnras \textbf{217}, 805
  (1985).

\bibitem{HS85} Y.~Hoffman and J.~Shaham, \apj \textbf{297}, 16 (1985).

\bibitem{OPW86} S.~Otto, H.~D.~Politzer, and M.~B.~Wise, \prl
  \textbf{56}, 1878 (1986).

\bibitem{Cou87} H.~M.~P.~Couchman, \mnras \textbf{225}, 777 (1987).

\bibitem{Col89} P.~Coles, \mnras \textbf{238}, 319 (1989).

\bibitem{LHP89} S.~L.~Lumsden, A.~F.~Heavens, and J.~A.~Peacock,
  \mnras \textbf{238}, 293 (1989).

\bibitem{PH90} J.~A.~Peacock and A.~F.~Heavens, \mnras \textbf{243},
  133 (1990).

\bibitem{RS95} E.~Reg\H{o}s and A.~S.~Szalay, \mnras \textbf{272}, 447
  (1995).

\bibitem{Des08} V.~Desjacques, \prd \textbf{78}, 103503 (2008).

\bibitem{MC19} T.~Matsubara and S.~Codis, arXiv:1910.09561
  [astro-ph.CO]. 
  
\bibitem{DJS18} 
  V.~Desjacques, D.~Jeong and F.~Schmidt,
  \physrep  \textbf{733}, 1 (2018).

\bibitem{BKMR04} 
  N.~Bartolo, E.~Komatsu, S.~Matarrese and A.~Riotto,
  \physrep \textbf{402}, 103 (2004).
  
\bibitem{BCGS02} F.~Bernardeau, S.~Colombi, E.~Gazta\~naga and
  R.~Scoccimarro, \physrep \textbf{367}, 1 (2002).


\bibitem{JW99} B.~Jain and L.~V.~Van Waerbeke, \apj \textbf{530}, L1
  (2000).

\bibitem{HTY04} T.~Hamana, M.~Takada and N.~Yoshida, \mnras
  \textbf{350}, 893 (2004). 

\bibitem{MSB09} 
  L.~Marian, R.~E.~Smith and G.~M.~Bernstein,
  \apj \textbf{698}, L33 (2009).

\bibitem{MAPB10} M.~Maturi, C.~Angrick, F.~Pace and M.~Bartelmann,
  \aap \textbf{519}, A23 (2010). 

\bibitem{FSL10} Z.~Fan, H.~Shan and J.~Liu,\apj \textbf{719}, 1408
  (2010). 
  
\bibitem{Yan11} X.~Yang J.~M.~Kratochvil, S.~Wang E.~A.~Lim, Z.~Haiman
  and M.~May, \prd \textbf{84}, 043529 (2011).

\bibitem{Mar11} 
  L.~Marian, S.~Hilbert, R.~E.~Smith, P.~Schneider and V.~Desjacques,
  \apj \textbf{728}, L13 (2011).
  
\bibitem{MSHS13} 
  L.~Marian, R.~E.~Smith, S.~Hilbert and P.~Schneider,
  \mnras \textbf{432}, 1338 (2013).

\bibitem{Liu14} 
  J.~Liu, A.~Petri, Z.~Haiman, L.~Hui, J.~M.~Kratochvil and M.~May,
  \prd \textbf{91}, 063507 (2015).

\bibitem{Liu15} 
  X.~Liu {\it et al.},
  \mnras \textbf{450}, 2888 (2015).

\bibitem{LK15a} 
  C.~A.~Lin and M.~Kilbinger,
  \aap \textbf{576}, A24 (2015).

\bibitem{HSKM15} T.~Hamana, J.~Sakurai, M.~Koike and L.~Miller, \pasj
\textbf{67}, 34 (2015).

\bibitem{LK15b} 
  C.~A.~Lin and M.~Kilbinger,
  \aap \textbf{583}, A70 (2015).

\bibitem{OSY15}
K.~Osato, M.~Shirasaki and N.~Yoshida, \apj \textbf{806}, 186 (2015). 
  
\bibitem{Kac16} 
  T.~Kacprzak {\it et al.} [DES Collaboration],
  \mnras \textbf{463}, 3653 (2016).

\bibitem{Pee18} 
  A.~Peel, V.~Pettorino, C.~Giocoli, J.~L.~Starck and M.~Baldi,
  \aap \textbf{619}, A38 (2018).

\bibitem{Sha18} 
  H.~Shan {\it et al.},
  \mnras \textbf{474}, 1116 (2018).

\bibitem{Mar18} 
  N.~Martinet {\it et al.},
  \mnras \textbf{474}, 712 (2018).

\bibitem{Li19} 
  Z.~Li, J.~Liu, J.~M.~Z.~Matilla and W.~R.~Coulton,
  \prd \textbf{99}, 063527 (2019).

\bibitem{Cou19} W.~R.~Coulton, J.~Liu, I.~G.~McCarthy and K.~Osato,
  arXiv:1910.04171 [astrho-ph.CO]

\bibitem{ZN67} Y.~B.~Zel’dovich and I.~D.~Novikov, Sov.~Astron.
  \textbf{10}, 602 (1967).

\bibitem{Haw71} S.~Hawking, \mnras \textbf{152}, 75 (1971).

\bibitem{CH74} B.~J.~Carr and S.~W.~Hawking, \mnras \textbf{168}, 399
  (1974).
  
\bibitem{CKSY10} B.~J.~Carr, K.~Kohri, Y.~Sendouda and J.~Yokoyama,
  \prd \textbf{81}, 104019 (2010).

\bibitem{GLMS04} 
  A.~M.~Green, A.~R.~Liddle, K.~A.~Malik and M.~Sasaki,
  \prd \textbf{70}, 041502 (2004).

\bibitem{YHGK18} C.-M.~Yoo, T.~Harada, J.~Garriga and K.~Kohri, \ptep
  \textbf{2018}, 123E01 (2018).

\bibitem{GM19} C.~Germani and I.~Musco
\prl \textbf{122}, 141302  (2019).

\bibitem{DeL19} 
  V.~De Luca, G.~Franciolini, A.~Kehagias, M.~Peloso, A.~Riotto and C.~Ünal,
  \jcap \textbf{1907}, 048 (2019).

\bibitem{SY19} 
  T.~Suyama and S.~Yokoyama,
  arXiv:1912.04687 [astro-ph.CO].

\bibitem{GS19} 
  C.~Germani and R.~K.~Sheth,
  arXiv:1912.07072 [astro-ph.CO].

\bibitem{Blo16} J.~K.~Bloomfield, S.~H.~P.~Face, A.~H.~Guth, S.~Kalia,
  C.~Lam and Z.~Moss, arXiv:1612.03890 [math-ph].

\bibitem{Blo18} J.~K.~Bloomfield, S.~H.~P.~Face, A.~H.~Guth, S.~Kalia
  and Z.~Moss, arXiv:1810.02078 [math-ph].

\bibitem{PGP09} D.~Pogosyan, C.~Gay and C.~Pichon, \prd \textbf{80}, 081301
  (2009); \prd \textbf{81}, 129901(E) (2010). 

\bibitem{GPP12} C.~Gay, C.~Pichon and D.~Pogosyan, \prd \textbf{85},
  023011 (2012).

\bibitem{Mat94} T.~Matsubara, \apj \textbf{434}, L43 (1994).

\bibitem{Mat03} T.~Matsubara, \apj \textbf{584}, 1 (2003).

\bibitem{Cha67} J.~M.~Chambers, Biometrika \textbf{54}, 367 (1967).

\bibitem{Jus95} R.~Juszkiewicz, D.~H.~Weinberg, P.~Amsterdamski,
  M.~Chodorowski and F.~Bouchet, \apj \textbf{442}, 39 (1995).
  
\bibitem{Mat95} T.~Matsubara, \apjs \textbf{101}, 1 (1995).

\bibitem{Ame96} L.~Amendola,  \mnras \textbf{283}, 983 (1996).
  
\bibitem{DGR13} 
  V.~Desjacques, J.~O.~Gong and A.~Riotto,
  \jcap \textbf{1309}, 006 (2013).



\bibitem{CLM88} P.~Catelan, F.~Lucchin, S.~Matarrese, \prl
  \textbf{61}, 267 (1988)

\bibitem{GW86} B.~Grinstein and M.~B.~Wise, \apj \textbf{310}, 19
  (1986).

\bibitem{MLB86} S.~Matarrese, F.~Lucchin and S.~A.~Bonometto, \apjl
  \textbf{310}, L21 (1986).
  
\bibitem{ARW06} A.~P.~A.~Andrade, A.~L.~B.~Ribeiro and C.~A.~Wuensche,
  \aap \textbf{457}, 385 (2006).

  
\bibitem{LMD16} T.~Lazeyras, M.~Musso and V.~Desjacques, \prd
\textbf{93}, 063007 (2016).
  
\bibitem{MD16} T.~Matsubara and V.~Desjacques, \prd \textbf{93},
  123522 (2016).
  





\bibitem{Diz16} A.~Moradinezhad~Dizgah, K.~C.~Chan, J.~Nore{\~n}a,
  M.~Biagetti and V.~Desjacques, \jcap \textbf{1609}, 030 (2016).

\bibitem{BE87} J.~R.~Bond and G.~Efstathiou, \mnras \textbf{226}, 655
  (1987). 
    
\bibitem{Bal13} T.~Baldauf, U.~Seljak, R.~E.~Smith, N.~Hamaus and
  V.~Desjacques, \prd \textbf{88}, 083507 (2013).

\bibitem{Bal16} T.~Baldauf, S.~Codis, V.~Desjacques and C.~Pichon, \mnras
  \textbf{456}, 3985 (2016).

\bibitem{CPP18} S.~Codis, D.~Pogosyan and C.~Pichon, \mnras
  \textbf{479}, 973 (2018).

\bibitem{SVM16} M.~Schmittfull, Z.~Vlah, P.~McDonald, \prd \textbf{93},
  103528 (2016)

\bibitem{SV16} M.~Schmittfull, Z.~Vlah, 2016, \prd \textbf{94}, 103530
  (2016)

\bibitem{MFHB16} J.~E.~McEwen, X.~Fang, C.~M.~Hirata and J.~A.~Blazek,
  \jcap, \textbf{9}, 015 (2016)

\bibitem{FBMH17} X.~Fang, J.~A.~Blazek, J.~E.~McEwen and C.~M.~Hirata,
\jcap, \textbf{2}, 030 (2017)

\bibitem{Ham00} A.~J.~S.~Hamilton, \mnras \textbf{312}, 257 (2000).
  
\bibitem{class11} J.~ Lesgourgues, arXiv:1104.2932.
  
\bibitem{CLASS} D. Blas, J. Lesgourgues, T. Tram, \jcap \textbf{7} 034
  (2011).

\bibitem{Planck2018} Planck Collaboration, arXiv:1807.06209 (2018).

\bibitem{Lim53} D.~N.~Limber, \apj \textbf{117}, 134 (1953).
  
\bibitem{Kai98} N.~Kaiser, \apj \textbf{498}, 26 (1998).
  

\bibitem{Smi03} R.~E.~Smith, J.~A.~Peacock, A.~Jenkins, et al., \mnras
  \textbf{341}, 1311 (2003).

\bibitem{Tak12} 
  R.~Takahashi, M.~Sato, T.~Nishimichi, A.~Taruya and M.~Oguri,
  \apj  {\bf 761}, 152 (2012).

\bibitem{Laz15} 
  A.~Lazanu, T.~Giannantonio, M.~Schmittfull and E.~P.~S.~Shellard,
  \prd {\bf 93}, 083517 (2016).

\bibitem{Bos19} 
  B.~Bose, J.~Byun, F.~Lacasa, A.~Moradinezhad Dizgah and L.~Lombriser,
  arXiv:1909.02504 [astro-ph.CO].

\bibitem{Tak19} 
  R.~Takahashi, T.~Nishimichi, T.~Namikawa, A.~Taruya, I.~Kayo,
  K.~Osato, Y.~Kobayashi and M.~Shirasaki, 
  arXiv:1911.07886 [astro-ph.CO].

\bibitem{DCSS10} V.~Desjacques, M.~Crocce, R.~Scoccimarro and
  R.~K.~Sheth, \prd \textbf{82}, 103529 (2010).

\bibitem{Mat12} T.~Matsubara, \prd \textbf{86}, 063518 (2012).

\bibitem{Ma85} S.-K.~Ma, Statistical Mechanics (Philadelphia; World
  Scientific, 1985)

\bibitem{SB91} R.~J.~Scherrer and E.~Bertschinger, \apj \textbf{381},
  349 (1991).

\bibitem{BK95} F.~Bernardeau and L.~Kofman, \apj \textbf{443}, 479
  (1995). 
  
\end{thebibliography}

\twocolumngrid

\end{document}